\documentclass[lettersize,journal]{IEEEtran}
\usepackage{amsmath,amsfonts}
\usepackage{algorithmic}
\usepackage{algorithm}
\usepackage{array}
\usepackage[caption=false,font=normalsize,labelfont=sf,textfont=sf]{subfig}
\usepackage{textcomp}
\usepackage{stfloats}
\usepackage{url}
\usepackage{verbatim}
\usepackage{graphicx}
\usepackage{cite}
\usepackage{multirow}
\usepackage{makecell}
\usepackage{amsfonts}
\usepackage{diagbox}
\usepackage{color}
\usepackage[]{pifont}
\usepackage{amsmath}

\usepackage{tikz}

\begin{document}

\title{OPAF: Optimized Secure Two-Party Computation Protocols for Nonlinear Activation Functions in Recurrent Neural Network}

\author{
  Qian~Feng,~
  Zhihua~Xia,~\IEEEmembership{Member,~IEEE,}
  Zhifeng~Xu,~
  Jiasi~Weng,~
  and ~Jian~Weng,~\IEEEmembership{Member,~IEEE}
\thanks{
  Qian Feng, Zhihua Xia, Zhifeng Xu, Jiasi Weng and Jian Weng are with the Department of College of Cyberspace Security, Jinan University, Guangzhou, 510632, China.

%
%


  Zhihua Xia is the corresponding author. e-mail: xia\_zhihua@163.com.
  \protect\\
  }
}

\markboth{Journal of \LaTeX\ Class Files,~Vol.~14, No.~8, August~2021}%
{Shell \MakeLowercase{\textit{et al.}}: A Sample Article Using IEEEtran.cls for IEEE Journals}


\maketitle

\begin{abstract}
 Deep neural network (DNN) typically involves convolutions, pooling, and activation function. Due to the growing concern about privacy, privacy-preserving DNN becomes a hot research topic. Generally, the convolution and pooling operations can be supported by additive homomorphic and secure comparison, but the secure implementation of activation functions is not so straightforward for the requirements of accuracy and efficiency, especially for the non-linear ones such as exponential, \textsf{sigmoid}, and \textsf{tanh} functions. This paper pays a special attention to the implementation of such non-linear functions in semi-honest model with two-party settings, for which SIRNN is the current state-of-the-art. Different from previous works, we proposed improved implementations for these functions by using their intrinsic features as well as worthy tiny tricks. At first, we propose a novel and efficient protocol for exponential function by using a divide-and-conquer strategy with most of the computations executed locally. Exponential protocol is widely used in machine learning tasks such as Poisson regression, and is also a key component of sigmoid and tanh functions. Next, we take advantage of the symmetry of \textsf{sigmoid} and \textsf{Tanh}, and fine-tune the inputs to reduce the 2PC building blocks, which helps to save overhead and improve performance. As a result, we implement these functions with fewer fundamental building blocks. The comprehensive evaluations show that our protocols achieve state-of-the-art precision while reducing run-time by approximately $57\%$, $44\%$, and $42\%$ for exponential (with only negative inputs), sigmoid, and Tanh functions, respectively.
\end{abstract}


\section{Introduction}
\IEEEPARstart{A}{rtificial} Intelligence (AI) is rapidly evolving in recent years and being used in a variety of fields, including health care, finance, manufacturing, and beyond. AI requires large amounts of data for learning, and as AI continues to evolve, numerous privacy problems related to it aroused widespread concern. European Union has set up the General Data Privacy Regulation~\cite{voigt2017eu}, which specifies that the utilization of personal data requires the consent of the data subject. Also it has motivated the development of privacy-preserving DNN training and inference.

  The seminal work of SecureML~\cite{mohassel2017secureml} demonstrated that privacy-preserving inference and training of DNN can be resolved by secure two-party computation (2PC). 2PC~\cite{goldreich2019play,yao1982protocols} allows the two parties, $P_1$ and $P_2$, to interactively compute an agreed function $f$ on their sensitive inputs $x$ and $y$, with strong guarantees that the interaction discloses no information about the sensitive inputs. There has recently been a surge of 2PC-based works for privacy-preserving DNN inference~\cite{liu2017oblivious,feng2020securenlp,rathee2020cryptflow2,mishra2020delphi,patra2021aby2,rathee2021sirnn,huang2022cheetah,gupta2022llama}.
  Despite the awesome efforts, there remains a significant gap from privacy-preserving machine learning to practical applications for performance. Also, the 2PC secure inference protocols that support complex neural networks effectively and efficiently stand as an open problem. As exhibited in Table~\ref{tab:related1}, a multitude of existing works are suffering from a wide range of deficiencies.  CrypTFlow2~\cite{rathee2020cryptflow2} and Cheetah~\cite{huang2022cheetah} have achieved great success in privacy-preserving CNN inference (e.g., ResNet~\cite{he2016deep}, DenseNet ~\cite{huang2017densely}, and MobileNet ~\cite{sandler2018mobilenetv2}) through 2PC. However, these networks only involve simple nonlinear functions such as ReLU and Maxpool.

  With the rising of ChatGPT, a popular language model that generates human-like responses in natural language conversations, the spotlight has once again been turned towards Recurrent Neural Networks (RNNs). RNNs are neural networks that can process sequential or time series data and are commonly used for natural language processing tasks such as speech recognition and machine translation. These networks extensively use nonlinear activation functions that are more complex, such as exponential, \textsf{sigmoid}, and \textsf{Tanh} functions. Unfortunately only a few works have provided secure implementations of these nonlinear functions~\cite{mohassel2017secureml,liu2017oblivious,keller2020mp}, and those works are still not satisfactory.

\begin{table*}[!t]
\caption{Some related works. RNN represents whether the work supports recurrent neural network structure and \ding{73} represents the baseline.}
\label{tab:related1}
\centering
\setlength{\tabcolsep}{3.2mm}{
\begin{tabular}{|c|c|ccc|ccc|ccc|}
\hline
\multirow{2}{*}{\textsf{Work}} & \multirow{2}{*}{\textsf{RNN}} & \multicolumn{3}{c|}{\textsf{Run-time}}                                & \multicolumn{3}{c|}{\textsf{Communication traffic}}
& \multicolumn{3}{c|}{\textsf{Max Errors}}
\\ \cline{3-11}&
& \multicolumn{1}{c|}{\textsf{sigmoid}}
& \multicolumn{1}{c|}{\textsf{Tanh}}
& \textsf{exp}
& \multicolumn{1}{c|}{\textsf{sigmoid}}
& \multicolumn{1}{c|}{\textsf{Tanh}}
& \textsf{exp}
& \multicolumn{1}{l|}{\textsf{sigmoid}}
& \multicolumn{1}{c|}{\textsf{Tanh}}
& \textsf{exp}
\\ \hline
\textsf{SIRNN~\cite{rathee2021sirnn}}
& \ding{51}
& \multicolumn{1}{c|}{\ding{73}}
& \multicolumn{1}{c|}{\ding{73}}
& \ding{73}
& \multicolumn{1}{c|}{\ding{73}}
& \multicolumn{1}{c|}{\ding{73}}
& \ding{73}
& \multicolumn{1}{c|}{\ding{73}}
& \multicolumn{1}{c|}{\ding{73}}
& \ding{73}    \\ \hline
\textsf{\makecell[c]{MiniONN~\cite{liu2017oblivious}\\12-piece}}
& \ding{51}
& \multicolumn{1}{c|}{\ding{73}$\times34$}
& \multicolumn{1}{c|}{-}
& -
& \multicolumn{1}{c|}{\ding{73}$\times19$}
& \multicolumn{1}{c|}{-}
& -
& \multicolumn{1}{c|}{\ding{73}$\times34$}
& \multicolumn{1}{c|}{-}
& -   \\ \hline
\textsf{\makecell[c]{MiniONN~\cite{liu2017oblivious}\\48-piece}}
& \ding{51}
& \multicolumn{1}{c|}{\ding{73}$\times115$}
& \multicolumn{1}{c|}{-}
& -
& \multicolumn{1}{c|}{\ding{73}$\times70$}
& \multicolumn{1}{c|}{-}
& -
& \multicolumn{1}{c|}{\ding{73}}
& \multicolumn{1}{c|}{-}
& - \\ \hline

\textsf{MP-SPDZ~\cite{keller2020mp}}
& -
& \multicolumn{1}{c|}{\ding{73}$\times89$}
& \multicolumn{1}{c|}{-}
& \ding{73}$\times70$
& \multicolumn{1}{c|}{\ding{73}$\times201$}
& \multicolumn{1}{c|}{-}
& \ding{73}$\times70$
& \multicolumn{1}{c|}{\ding{73}}
& \multicolumn{1}{c|}{-}
& \ding{73}$\times70$  \\ \hline
\textsf{CryptFlow2~\cite{rathee2020cryptflow2}}
& \ding{55}
& \multicolumn{1}{c|}{\ding{55}}
& \multicolumn{1}{c|}{\ding{55}}
& \ding{55}
& \multicolumn{1}{c|}{\ding{55}}
& \multicolumn{1}{c|}{\ding{55}}
& \ding{55}
& \multicolumn{1}{c|}{\ding{55}}
& \multicolumn{1}{c|}{\ding{55}}
& - \\ \hline
\textsf{Cheetah~\cite{huang2022cheetah}}
& \ding{55}
& \multicolumn{1}{c|}{\ding{55}}
& \multicolumn{1}{c|}{\ding{55}}
& \ding{55}
& \multicolumn{1}{c|}{\ding{55}}
& \multicolumn{1}{c|}{\ding{55}}
& \ding{55}
& \multicolumn{1}{c|}{\ding{55}}
& \multicolumn{1}{c|}{\ding{55}}
& \ding{55}  \\ \hline
\textsf{Ours}
& \ding{51}
& \multicolumn{1}{c|}{\ding{73}$\times56\%$}
& \multicolumn{1}{c|}{\ding{73}$\times58\%$}
& \ding{73}$\times43\%$
& \multicolumn{1}{c|}{\ding{73}$\times64\%$}
& \multicolumn{1}{c|}{\ding{73}$\times60\%$}
& \ding{73}$\times59\%$
& \multicolumn{1}{c|}{\ding{73}$\times33\%$}
& \multicolumn{1}{c|}{\ding{73}}
& \ding{73}    \\ \hline
\end{tabular}
}
\end{table*}

  The demands above drive us to further optimize the 2PC protocols for nonlinear activation functions. The contributions of this paper can be summarized as follows:

\begin{itemize}
  \item We propose a novel 2PC implementation for exponential function using a divide-and-conquer strategy with most of the computations executed in local. The proposed protocol requires only a constant number of rounds and low communication in high-precision fixed-point operations. It provides an efficiency building block for the 2PC implementation of  \textsf{sigmoid} and \textsf{Tanh}. 
  \item We improve the existing realizations of \textsf{sigmoid} and \textsf{Tanh} from ~\cite{rathee2021sirnn} by reducing 2PC building blocks through the symmetry of functions. In addition, we further reduce the building blocks by fine-tuning the encoding of $-x$ (with $x\ge 0)$. These save cost and runtime in both respects.
  \item Finally, we conducted a comprehensive evaluation to our secure exponential, \textsf{sigmoid}, and \textsf{Tanh} functions. In addition, we also apply our secure \textsf{sigmoid} and \textsf{Tanh} protocol to an entire process of RNN inference, which further demonstrates our improvements. 
\end{itemize}

\section{Related Works}
  Secure Multiparty Computation (MPC) is a computer security technology that allows multiple participants to jointly compute a function without revealing their inputs to each other, only disclosing the computed result. MPC is often modeled as a Boolean circuit~\cite{micali1987play,yao1982protocols,rouhani2018deepsecure} or an arithmetic circuit~\cite{mohassel2017secureml,liu2017oblivious,keller2020mp,patra2021aby2,rathee2020cryptflow2,huang2022cheetah,fan2022nfgen,rathee2021sirnn}, and can be implemented through two different routes: the secret sharing and the garbled circuit routes. The secret sharing route requires interaction among the parties for each nonlinear gate of the circuit, resulting in lower communication bandwidth but with the loop count linearly related to the circuit depth. The garbled circuit route involves each party constructing an encrypted version of the circuit, allowing only one computation and resulting in higher communication bandwidth but a constant number of rounds. In the past, secret sharing was more suitable for linear operations, while garbled circuits were more suitable for nonlinear operations. ABY~\cite{demmler2015aby} proposed the conversion between Boolean and arithmetic circuits to improve overall performance.

  In recent years, with the rapid development of privacy-preserving machine learning, arithmetic secret sharing-based semi-honest 2PC non-linear computation has achieved a historic breakthrough~\cite{mohassel2017secureml, liu2017oblivious, keller2020mp, patra2021aby2, rathee2020cryptflow2, huang2022cheetah, fan2022nfgen, rathee2021sirnn}. The performance of many non-linear activation functions widely used in neural networks has surpassed that of garbled circuits in secret sharing methods.

  Non-linear function computation based on arithmetic secret sharing is mainly divided into two technical routes: (1) ad hoc piecewise linear approximation route represented in SecureML~\cite{mohassel2017secureml}, ABY2.0~\cite{patra2021aby2} and MiniONN~\cite{liu2017oblivious}, and (2) general MPC routes based on protocol combinations, which is represented by MP-SPDZ~\cite{keller2020mp} and SIRNN~\cite{rathee2021sirnn}. The use of piecewise linear approximations requires developer intervention for each dataset and each model to balance accuracy and latency, which might be unacceptable in the context of automated frameworks for secure inference. Additionally, these methods rely on 2PC building blocks from~\cite{demmler2015aby, keller2020mp, yao1986generate} and suffer from huge performance overheads. SecureML \cite{mohassel2017secureml} and ABY2.0~\cite{patra2021aby2} use a simple three-piece linear approximation of sigmoid. This simple implementation has a whopping error that greatly affects the accuracy of inference and training. MiniONN~\cite{liu2017oblivious} indicated that this approximation causes the cross-entropy loss to diverge to infinity. It uses a 12-piece spline approximation, and achieves smaller errors than SecureML and ABY2.0 at a higher cost, so as to meet the requirements of cross-entropy loss. Recently, Fan et al.~\cite{fan2022nfgen} propose a non-linear function code generator (NFGen) for evaluating nonlinear functions on a general MPC platform. They use $m$-piecewise polynomials with a maximum degree of $k$ to approximate each non-linear function, and propose an algorithm for automatically determining $k$ and $m$. Their scheme results in positive improvements on some nonlinear functions, however, it achieves no improvement on \textsf{sigmoid} compared to~\cite{keller2020mp}.

  As a general MPC protocol, ~\cite{keller2020mp} computes $\textsf{sigmoid}=\frac{1}{1+e^{-x}}$ in three steps. First, the exponent is approximated with a Taylor series polynomial. Then, $1+e^{-x}$ in secret sharing can be computed locally. Finally, the result of $\textsf{sigmoid}$ is obtained by using a secure computation protocol for computing the reciprocal of $1+e^{-x}$. Most existing secure reciprocal computation algorithms are based on iterative algorithms such as Newton-Raphson's method~\cite{wagh2021f, ge2021practical} and Goldschmidt's algorithms~\cite{goldschmidt1964applications, ito1997efficient}. These algorithms can securely compute the reciprocal of a number $v$. If $\alpha$ is known such that $2^{\alpha}\le v < 2^{\alpha+1}$, the iteration process can be started directly, otherwise $\alpha$ needs to be securely computed at first, which incurs huge overhead. In addition, ~\cite{keller2020mp} requires high-degree Taylor series polynomials to accurately approximate the exponential function, which also incurs huge overhead.

  The state-of-the-art 2PC work, SIRNN~\cite{rathee2021sirnn}, revealed that $\alpha$ corresponding to $1+e^{-x}$ is always 0 when $x>0$; thus $\alpha$ does not need to be computed securely and the reciprocal of $1+e^{-x}$ can be directly calculated. Similarly, when $x<0$, we have $\textsf{sigmoid}(x)=e^x\cdot\frac{1}{1+e^{x}}$ and the $\alpha$ corresponding to $1+e^{x}$ is always 0, in which it is also not needed to compute $\alpha$. Finally, it is obvious that $\textsf{sigmoid}(x) = 0.5$ for $x = 0$. Based on this finding, SIRNN represents the \textsf{sigmoid} function as a piecewise function:

\[\textsf{sigmoid}(x)=\begin{cases}
0.5,&if\ x=0,\\
\frac{1}{1+e^{-x}},&if\ x>0, \\
e^x\cdot\frac{1}{1+e^{x}},&if\ x<0,
\end{cases}\]

\noindent which avoids computing $\alpha$ in secure multiparty scenarios and greatly reduces the overhead of \textsf{sigmoid}. The \textsf{Tanh} function and the \textsf{sigmoid} function are closely related~\cite{goodfellow2016deep}: $\textsf{Tanh}(x)=2\cdot\textsf{sigmoid}(2x)- 1$. Therefore, the 2PC implementation of \textsf{Tanh} can be achieved by secure \textsf{sigmoid}. However, as pointed out in~\cite{gupta2022llama}, the functionalities provided in SIRNN are highly sequential and would lead to a large number of rounds in the online phase. Nevertheless, SIRNN is still the state-of-the-art work for 2PC \textsf{sigmoid} and \textsf{Tanh}. In this work we significantly reduced the number of rounds and online phases of SIRNN.

SIRNN also proposed a secure protocol for the exponential function that receives negative inputs by using lookup tables~\cite{dessouky2017pushing}. When the length of input is small (e.g., $\leq$8 bits), this method can achieve excellent efficiency and accuracy. However, the cost of lookup tables grows exponentially with the length of input. To overcome this obstacle, SIRNN has made some optimizations to their exponential protocol. Specifically, a long secret input is securely decomposed into multiple short secret inputs. Then, the exponential results of multiple short inputs are calculated separately, and all results from the previous step are multiplied together using a multiplication protocol to obtain the final output. While these optimizations somewhat suppress the problem of the overhead explosion, the secure decomposition of secret inputs and multiplication operations still introduce a significant overhead which cannot be ignored.

 \textbf{Summary}. Only a few works have provided 2PC implementations for non-linear functions such as exponential, \textsf{sigmoid} and \textsf{Tanh}. Among them, SIRNN's implementations ~\cite{rathee2021sirnn} are state-of-the-art currently available. However, their 2PC implementation for exponential is based on a lookup table route with extensive communications. And their 2PC implementations for \textsf{sigmoid} and \textsf{Tanh} fail to take the advantage of the symmetry in these functions to save more costs (more details are available in Section~\ref{sec:Sigmoid}).

\section{Preliminaries}
We define the expected notations, the 2PC building blocks from existing works, and the threat model we against. As with most practical MPC platforms, we use \textit{fixed-point} arithmetics instead of the commonly used \textit{floating-point}  arithmetics for efficiency. Accordingly, it is crucial to first encode the \textit{floating-point} or $real$ number in the plaintext environment into its corresponding \textit{fixed-point} representations. Our encoding details are defined in Section~\ref{sec:Fixed-Point}. Second, it is extremely vital to accurately evaluate the gap between the results on fixed-points and those on floating-points. In this paper, we use $ULP$ (units in last place) errors to report the precision of our implementations. The details of the ulp error are defined in Section~\ref{sec:ULP}. Additionally, our implementation is based on the secret sharing technique, and it is defined in Section~\ref{sec:SS}. Our implementations are defend against static semi-honest adversaries running in a probabilistic polynomial (PPT), which we present in Section~\ref{sec:ThreatModel} and they use 2PC building blocks from existing work in Section~\ref{sec:Func}.
Finally, let $\lambda$ denote the computational security parameter, which is set to 128 by default.

\subsection{Fixed-Point Representation}\label{sec:Fixed-Point}

In \textit{fixed-point} arithmetic, a float number $x_f \in \mathbb{R}$ is (approximately) represented using an $l$-bit integer $x$ as,

\begin{equation}
\label{equ:Fix}
\begin{aligned}
x&=\textsf{Fix}(x_f,s)\\
&=\lfloor x_f \cdot 2^s\rfloor \ mod \  L,
\end{aligned}
\end{equation}

\noindent where $L=2^l$, $x \in \mathbb{Z}_L$, $l$ denotes the bitwidth, and $s \in \mathbb{Z}$ is the scale indicating the fractional part of bitwidth. For easy representation, we define an indicator function $1\{b\}$ that returns 1 when $b$ is true, and 0 otherwise. Furthermore, we define $\textsf{msb}(x)$ to calculate the most important bit of a fixed-point number $x$ as

\begin{equation}
\label{equ:msb}
\begin{aligned}
    \textsf{msb}(x)=1\{x \geq 2^{l-1}\},
\end{aligned}
\end{equation}

\noindent define $\textsf{sfp}(x)$ to transform the number $x\in \mathbb{Z}_L$ to be a signed-fixed-point number as

\begin{equation}
\label{equ:sfp}
\begin{aligned}
    \textsf{sfp}(x)=x-\textsf{msb}(x)\cdot L.
\end{aligned}
\end{equation}

\noindent Then, the reverse mapping from fixed-point number $x$ to float-point representation can be defined as
\begin{equation}
\label{equ:rev}
\begin{aligned}
x'_f&=\textsf{rev}(x),\\
    &=\textsf{sfp}(x)/2^s,
\end{aligned}
\end{equation}
where the division is operated over $\mathbb{R}$.

\subsection{ULP Errors}\label{sec:ULP}
It is impractical to precisely represent an irrational value using a finite number of bits. As a result, finding a way to quantify the biases between an exact real result and the output of a math library with a finite bit representation is critical. Various error concepts can be used -- absolute error, relative error and ULP error. Standard math libraries utilize ULP error as a standard to evaluate whether the real output of a math function is close enough to the finite-bit output produced by the library~\cite{wang2014intel}. The precision and accuracy of the realization of that math function increase as the ULP value decreases. At a high level, ULP error between an exact real result $r$ and the output $res$ of secure computation is equal to the number of representable numbers between $r$ and $res$ ~\cite{rathee2021sirnn}.

\subsection{Secret Sharing}\label{sec:SS}

We use 2-out-of-2 secret sharing protocol over different power-of-2 rings to construct our new protocols as in~\cite{rathee2021sirnn}. Assuming there are two parties $P_1$ and $P_2$. For additive sharing, given a fixed-point number $x \in \mathbb{Z}_L$, it generates a random number denoted as $[x]_1 \in \mathbb{Z}_L$ for $P_1$, and then calculates $[x]_2=x-[x]_1~mod~L$ for $P_2$. We define a function $\textsf{wrap}([x]_1,[x]_2,L)$ to judge if $[x]_1+[x]_2$ is out of the range of $\mathbb{Z}_L$ as

\begin{equation}
\label{equ:msb}
\begin{aligned}
    \textsf{wrap}([x]_1,[x]_2,L)=1\{[x]_1+[x]_2 \ge L\}.
\end{aligned}
\end{equation}

\noindent Accordingly, a reconstruction of $x$ can be defined as

\begin{equation}
\label{equ:Reconst}
\begin{aligned}
    x   &=\textsf{Reconst}([x]_1,[x]_2)\\
        &=[x]_1+[x]_2 \ mod\ L, \\
        &=[x]_1+[x]_2-wrap(x)\cdot L.
\end{aligned}
\end{equation}

\noindent For boolean sharing, given a bit $x$, it generates a random bit denoted as $\left \langle x \right \rangle_1$ for $P_1$, and then calculates $\left \langle x \right \rangle_2 = \left \langle x \right \rangle_1 \oplus x$ for $P_2$. Accordingly, the reconstruction of $x$ can be defined as

\begin{equation}
\label{equ:Reconstb}
\begin{aligned}
    x   &=\textsf{Reconst}(\left \langle x \right \rangle_1,\left \langle x \right \rangle_2)\\
        &=\left \langle x \right \rangle_1 \oplus \left \langle x \right \rangle_2. \\
\end{aligned}
\end{equation}

\subsection{2PC and Threat Model}\label{sec:ThreatModel}
\noindent \textbf{2PC.} Secure 2-party computation (2PC)~\cite{goldreich2019play,yao1982protocols} allows two parties, $P_1$ and $P_2$, to compute an agreed function $f$ on their sensitive inputs $x$ and $y$. It provides an interactive protocol with strong guarantees that the interaction discloses no information about the sensitive inputs apart from what can be inferred from the output. A typical approach to 2PC starts with the parties secretly sharing their inputs with each other and running a protocol that uses the shares of ($x$, $y$) to securely generate the shares of $f(x, y)$. Then, the parties exchange the shares of the output and reconstruct the output.

\noindent \textbf{Threat Model.} Our threat model is the same as SIRNN~\cite{rathee2021sirnn} and considers a static semi-honest adversary running in probabilistic polynomial time (PPT). Briefly, there is a computationally constrained adversary $\mathcal{A}$ that takes over one of the parties at the beginning of the protocol execution, which conforms to the protocol specification but tries to learn additional information about the honest party's input. We argue for the security against this adversary in the standard simulation paradigm ~\cite{canetti2000security,goldreich2019play,lindell2017simulate}, which demonstrates the indistinguishability of the adversary's views in the real and ideal executions in the case where a trusted third party receives the inputs and provides the functional outputs alone. For a real function $f$ to be calculated, consider the following two interactions: real interaction, in which $P_1$ and $P_2$ interact using the protocol specification in the presence of $\mathcal{A}$ and the environment $\mathcal{Z}$, and ideal interaction, in which $P_1$ and $P_2$ send their inputs to a trusted function $\mathcal{F}$ that computes $f$ and return the outputs to parties. We believe that for every real adversary $\mathcal{A}$, there is an ideal adversary $\mathcal{S}$, so the environment $\mathcal{Z}$ interacting with the adversary cannot distinguish between real and ideal interactions. Our protocol calls for several sub-protocols. For ease of description, we use a hybrid model to describe them. This is the same as the actual interaction, except that the execution of the sub-protocols is replaced by the calls to the corresponding trusted functions - it is said that the protocol calling $\mathcal{F}$ is in the $\mathcal{F}$-hybrid model.

\subsection{2PC Functionalities}\label{sec:Func}
For a 2-party computation functionality $\mathcal{F}$, we say ``$P_1$ and $P_2$ invoke $\mathcal{F}(x)$ to learn $y$" to mean that $P_1$ with input $[x]_1$ and $P_2$ with input $[x]_2$ invoke $\mathcal{F}$ and learn additive shares of $y$, i.e., $P_1$ gets $[y]_1$ and $P_2$ gets $[y]_2$. In our
protocols, we use the following 2-party computation functionalities.

\begin{itemize}

  \item $\mathcal{F}_{\textsf{msb}}$ takes $[x]_i \in\mathbb{Z}_L$ as inputs, and outputs $\left \langle \textsf{msb}(x) \right \rangle_i \in \{0,1\}$, with $\textsf{msb}(x)= \textsf{Reconst}(\left \langle \textsf{msb}(x) \right \rangle_1$, $\left \langle \textsf{msb}(x) \right \rangle_2)$.
    It can be realized as in~\cite{rathee2021sirnn} with communication less than $\lambda l+14l$ bits.

  \item $\mathcal{F}_{\textsf{msbTOwrap}}$ takes $[x]_i\in\mathbb{Z}_L$ and $\left \langle \textsf{msb}(x) \right \rangle_i \in \{0,1\}$ as inputs, and outputs $\left \langle \textsf{wrap}(x) \right \rangle_i$, with $\textsf{wrap}(x)= \textsf{Reconst}(\left \langle \textsf{wrap}(x) \right \rangle_1$, $\left \langle \textsf{wrap}(x) \right \rangle_2)$. It can be realized by~\cite{rathee2021sirnn} with $\lambda+2$ bits of total communication.

  \item $\mathcal{F}_{\textsf{Mux}}$ takes $[x]_i \in \mathbb{Z}_L$ and $\left \langle b \right \rangle_i \in\{0,1\}$ as inputs, and outputs $[y]_i \in\mathbb{Z}_L$, with $y=(b)~?~x:0$. $\mathcal{F}_{\textsf{Mux}}$ can be realized as in~\cite{rathee2020cryptflow2} with $2(\lambda+ 2l)$ bits of total communication. \cite{rathee2021sirnn} provide an optimized protocol that reduces communication from $2(\lambda+ 2l)$ to $2(\lambda+ l)$.

  \item $\mathcal{F}_{\textsf{AND}}$ takes $\left \langle x \right \rangle_i \in \{0,1\}$ and $\left \langle y \right \rangle_i \in\{0,1\}$ as input, and returns $\left \langle z \right \rangle_i\in\{0,1\}$ with $z=x\&y$. It can be realized using Beaver bit-triples~\cite{beaver1992efficient, rathee2020cryptflow2} with $\lambda + 20 $ or 148 bits of total communication.

  \item $\mathcal{F}_{\textsf{CrossTerm}}$ takes $[x]_1 \in \mathbb{Z}_{2^m}$ and $[x]_2 \in \mathbb{Z}_{2^n}$ as inputs, and outputs $[y]_i\in\mathbb{Z}_L$ such that $y = [x]_1\times [x]_2$. It can be realized as in~\cite{rathee2021sirnn} with $\mu(\lambda+\mu/2+1/2)+mn$ bits of total communication where $\mu=\min(m,n)$.

  \item $\mathcal{F}_{\textsf{Mult}}$ takes $[x]_i \in \mathbb{Z}_{2^m}$ and $[y]_i \in \mathbb{Z}_{2^n}$ as inputs, and outputs $[z]_i\in\mathbb{Z}_L$  where $z = x \cdot y$ for $l=m+n$. It can be realized as in~\cite{rathee2021sirnn} with $\lambda(2\mu+6)+2\mu\nu+{\mu}^2+3\mu+2\nu+4$ bits of total communication where $\mu=\min(m,n),\nu=\max(m,n)$.

  \item $\mathcal{F}_{\textsf{B2A}}$ takes $\left \langle x \right \rangle_i$ as input and outputs arithmetic shares of the same value, i.e.,$[x]_i\in\mathbb{Z}_L$.  It can be realized as in~\cite{rathee2020cryptflow2} with $\lambda+l$ bits of total communication.

  \item $\mathcal{F}_{\textsf{TR}}$ takes $[x]_i \in \mathbb{Z}_L$ as inputs, and outputs $[y]_i \in \mathbb{Z}_{2^{l-s}}$ such that $y = \lfloor x / 2^s\rfloor$. $\mathcal{F}_{\textsf{TR}}$ can be realized as in~\cite{rathee2021sirnn} with $\lambda(s+1)+l+13s$ bits of total communication.

  \item $\mathcal{F}_{\textsf{Rec}}$ takes $[x]_i$ as input, and outputs $[\textsf{Fix}(\frac{1}{\textsf{rev}(x)}, s)]_i$. Note that $\mathcal{F}_{\textsf{Rec}}$ requires the input $\textsf{rev}(x) \in [1,2)$. $\mathcal{F}_{Rec}$ has been realized in ~\cite{rathee2021sirnn}.
\end{itemize}

\section{The Proposed 2PC Protocols}
In this section, we present our proposed exponential, sigmoid, and tanh protocols in detail, and analyze the security of the protocols. Our protocols are built on the basis of the existing 2PC functionalities described in subsection~\ref{sec:Func}.

\subsection{Exponential Function}\label{sec:Exponential}
The first protocol we proposed is a secure two-party computation for exponential function. It takes secret fixed-point numbers as input, and outputs secret fixed-point results for the function, which we denote as $\mathcal{F}_{\textsf{exp}}$. As explained in Section \ref{sec:SS}, we use 2-out-of-2 additive secret sharing. The secret input $x$ is split into two secret shares $[x]_1$ and $[x]_2$ held by $P_1$ and $P_2$, respectively. The intention of $P_i$ is to obtain the secret result of $\textsf{Fix}(e^{\textsf{rev}(x)}, s)$ without revealing any information about its $[{x}]_i$.

\noindent \textbf{Motivation}. SIRNN built a secure 2-PC protocol for exponential function by constructing the look-up tables for all items of $e^x, x \in \mathbb{Z}_L$. Then the required item is securely transmitted by oblivious transfer (OT). It achieves state-of-the art efficiency. Xia et al. proposed a secure 2-PC protocol for exponential function in infinite domain~\cite{xia2021str}. They designed the protocols to switch the additive and multiplicative shares, and used the following property of exponential function to construct the secure protocols,

\begin{equation}
\label{equ:expproperty}
    e^{a+b} = e^a \cdot e^b.
\end{equation}

\noindent The protocol in~\cite{xia2021str} is efficient but cannot be securely implemented in finite domain. Inspired by ~\cite{xia2021str}, we design a secure 2-PC protocol for exponential function using the property in Eq.(\ref{equ:expproperty}).

\noindent \textbf{Design}. Different from previous works, we designed an improved protocol to securely implement the exponential function by using divide-and-conquer strategy. Recalling Eqs.~(\ref{equ:sfp}), (\ref{equ:rev}), and (\ref{equ:Reconst}), we have
\begin{equation}
\label{equ:rev1}
\begin{aligned}
    x'_f    &=\textsf{rev}(x)=\textsf{sfp}(x)/2^s\\
            &=(x-\textsf{msb}(x)\cdot L)/{2^s}\\
            &=({[x]_1+[x]_2-\textsf{wrap}(x)\cdot L-\textsf{msb}(x) \cdot L})/{2^s}.
\end{aligned}
\end{equation}

We list all possible cases of $x'_f$ and $e^{x'_f}$ in Table~\ref{tab:sfpcase}, and deal with each case separately. For simplification, we write \textsf{wrap} and \textsf{msb} for \textsf{wrap}(x) and \textsf{msb}(x), respectively.

\begin{table}[!t]
\caption{The association from wrap and msb to $x'_f$ and $e^{x'_f}$.}
\label{tab:sfpcase}
\centering
\setlength{\tabcolsep}{4.6mm}{
\begin{tabular}{|c|c|l|l|}
\hline
\textsf{wrap} & \textsf{msb} & \multicolumn{1}{c|}{$x'_f$} & \multicolumn{1}{c|}{$e^{x'_f}$}  \\ \hline
1 & 1 & ~~$\frac{[x]_1+[x]_2-2L}{2^s}$ & ~~$e^{\frac{[x]_1+[x]_2-2L}{2^s}}$  \\ \hline
1 & 0 & ~~$\frac{[x]_1+[x]_2-L}{2^s}$ & ~~$e^{\frac{[x]_1+[x]_2-L}{2^s}}$  \\ \hline
0 & 1 & ~~$\frac{[x]_1+[x]_2-L}{2^s}$ & ~~$e^{\frac{[x]_1+[x]_2-L}{2^s}}$  \\ \hline
0 & 0 & ~~$\frac{[x]_1+[x]_2}{2^s}$ & ~~$e^{\frac{[x]_1+[x]_2}{2^s}}$  \\ \hline
\end{tabular}
}
\end{table}

Then, Eq.~(\ref{equ:rev1}) can be expressed as a segment function according to Table~\ref{tab:sfpcase},
\begin{equation}
\label{equ:revsegment0}
\begin{aligned}
x'_f=\textsf{rev}(x)&=\left \{
\begin{array}{ll}
    \frac{[x]_1+[x]_2-2L}{2^s}, & if\ \textsf{wrap}~\&~\textsf{msb}=1 \\ \\
    \frac{[x]_1+[x]_2-L}{2^s},  & if\ \textsf{wrap} \oplus \textsf{msb}=1 \\ \\
    \frac{[x]_1+[x]_2}{2^s},    & otherwise,
\end{array} \right.
\end{aligned}
\end{equation}
Then, we can calculate $e^{x'_f}$ as
\begin{equation}
\label{equ:Exponent}
\begin{aligned}
    e^{x'_f}&=e^{\frac{[x]_1+[x]_2-2L}{2^s}}\cdot(\textsf{wrap}\ \&\ \textsf{msb})\\
            &+e^{\frac{[x]_1+[x]_2-L}{2^s}}\cdot(\textsf{wrap} \oplus \textsf{msb})\\
            &+e^{\frac{[x]_1+[x]_2}{2^s}}\cdot(\textsf{wrap}\& \textsf{msb} \oplus \textsf{wrap} \oplus \textsf{msb} \oplus 1),
\end{aligned}
\end{equation}
which can be further transformed to be
\begin{equation}
\label{equ:ExponentOpt}
\begin{aligned}
    e^{x'_f}    &=(e^{\frac{[x]_1+[x]_2-2L}{2^s}}-e^{\frac{[x]_1+[x]_2}{2^s}}) \cdot (\textsf{wrap}~\&~\textsf{msb})\\
                &+(e^{\frac{[x]_1+[x]_2-L}{2^s}}-e^{\frac{[x]_1+[x]_2}{2^s}}) \cdot (\textsf{wrap} \oplus \textsf{msb})\\
                &+e^{\frac{[x]_1+[x]_2}{2^s}}.
\end{aligned}
\end{equation}

In 2PC scenario, $\textsf{msb}$ and $\textsf{wrap}$ can be calculated by $\mathcal{F}_{\textsf{msb}}$ and $\mathcal{F}_{\textsf{msbTOwrap}}$, operation $\&$ can be calculated by $\mathcal{F}_{\textsf{AND}}$, and operation $\oplus$ between boolean shares can be performed locally without any communication. And based on the mathematical properties of exponential functions, we next divide the operators in Eq.~(\ref{equ:Exponent}) into simpler operators, as follows:
\[
    \begin{cases}
        e^{\frac{[x]_1+[x]_2-2L}{2^s}}  &=e^{\frac{[x]_1-L}{2^s}} \times e^{\frac{[x]_2-L}{2^s}},\\
        e^{\frac{[x]_1+[x]_2-L}{2^s}}   &=e^{\frac{[x]_1-L/2}{2^s}} \times e^{\frac{[x]_2-L/2}{2^s}},\\
        e^{\frac{[x]_1+[x]_2}{2^s}}     &=e^{\frac{[x]_1}{2^s}} \times e^{\frac{[x]_2}{2^s}}.
\end{cases}\]

\noindent For the participants $P_1$ and $P_2$, since $P_1$ holds the secret share $[x]_1$, it can obtain $e^{\frac{[x]_1-L}{2^s}}$, $e^{\frac{[x]_1-L/2}{2^s}}$ and $e^{\frac{[x]_1}{2^s}}$ by local computation. Similarly, $P_2$ can obtain $e^{\frac{[x]_2-L}{2^s}}$, $e^{\frac{[x]_2-L/2}{2^s}}$ and $e^{\frac{[x]_2}{2^s}}$ through local computation by secret share $[x]_2$. Finally, the operator $\cdot$ in Eq.~(\ref{equ:Exponent}) can be calculated by $\mathcal{F}_{\textsf{Mux}}$ that allows the participants taking secret boolean shares and  arithmetic shares as input to output a new pair of secret arithmetic shares. Eq.~(\ref{equ:ExponentOpt}) is completely equivalent to Eq.~(\ref{equ:Exponent}), but it allows us to save one invocation of $\mathcal{F}_{\textsf{Mux}}$ in the realization. Our implementation details of $\mathcal{F}_{\textsf{exp}}$ are shown in Algorithm~\ref{alg:Exponential}.

\begin{algorithm}[t]
\caption{ Exponential protocol $\mathcal{F}_\textsf{exp}$}
\label{alg:Exponential}
\begin{algorithmic}[1]
\REQUIRE
   For $i \in \{1,2\}$, $P_i$ holds $[{x}]_i$.\\
\ENSURE
    For $i \in \{1,2\}$, $P_i$ learns $[{y}]_i$.

    \STATE $P_i$ sets:\\
    $\quad{a}_i     = \textsf{Fix}(e^{\frac{[x]_i}{2^s}},s')$,\\
    $\quad{b}_i     = \textsf{Fix}(e^{\frac{[x]_i-L/2}{2^s}},s')$,\\
    $\quad{c}_i     = \textsf{Fix}(e^{\frac{[x]_i-L}{2^s}},s')$ , where $s' > s$ and will be discussed in Section~\ref{sec:ULPerrors}.
    \STATE $P_i$ invokes $\mathcal{F}_{\textsf{CrossTerm}}$ with inputs $a_i$,$b_i$,$c_i$ to learn ${[a']_i}$,${[b']_i}$,${[c']_i}$ with $a'=a_1 \times a_2$, $b'=b_1 \times b_2$, and $c'=c_1 \times c_2$.
    \STATE $P_i$ invokes $\mathcal{F}_{\textsf{msb}}$ with input $[{x}]_i$ to learn output $\left \langle \textsf{msb} \right \rangle_i$.
    \STATE $P_i$ invokes $\mathcal{F}_{\textsf{msbTOwrap}}$ with inputs $\left \langle \textsf{msb} \right \rangle_i$ and $[{x}]_i$ to learn $\left \langle \textsf{wrap} \right \rangle_i$.
    \STATE $P_i$ invokes $\mathcal{F}_\textsf{AND}$ with inputs $\left \langle \textsf{wrap} \right \rangle _i$ and $\left \langle \textsf{msb}\right \rangle_i$ to learn output $\left \langle \textsf{wrap}\ \&\ \textsf{msb}\right \rangle_i$.
    \STATE $P_i$ sets:\\
     $\quad \left \langle {\textsf{wrap}\oplus \textsf{msb}}\right \rangle_i = \left \langle {\textsf{msb}}\right \rangle_i \oplus \left \langle {\textsf{wrap}}\right \rangle_i$,\\
    $\quad [b'-a']_i = [b']_i - [a']_i\ mod\ N$,\\
    $\quad [c'-a']_i = [c']_i - [a']_i\ mod\ N$, where $N \ge L$ and will be discussed in Section~\ref{sec:implementation}.
    \STATE $P_i$ invokes $\mathcal{F}_{\textsf{Mux}}$ with inputs $[b'-a']_i$ and $[{\textsf{wrap}\oplus \textsf{msb}}]_i$ to learn $[{t1}]_i$.
    \STATE $P_i$ invokes $\mathcal{F}_{\textsf{Mux}}$ with inputs $[c'-a']_i$ and $[{\textsf{wrap}~\&~ \textsf{msb}}]_i$ to learn $[{t2}]_i$.
    \STATE $P_i$ sets $[{rst}]_i = [{t1}]_i + [{t2}]_i + [a']_i\ mod\ N$.
    \STATE $P_i$ invokes $\mathcal{F}_{\textsf{TR}}$ with inputs $[{rst}]_i$ and common parameter $2s'-s$ to learn final output $[{y}]_i$.
\end{algorithmic}
\end{algorithm}
\noindent \textbf{Security analysis}. There are no interactions other than those generated by $\mathcal{F}_\textsf{CrossTerm}$, $\mathcal{F}_\textsf{msb}$, $\mathcal{F}_\textsf{msbTOwrap}$, $\mathcal{F}_\textsf{AND}$, $\mathcal{F}_\textsf{Mux}$, and $\mathcal{F}_{TR}$. The $[{y}]_i$ is randomly uniform and the security of $\mathcal{F}_\textsf{CrossTerm}$, $\mathcal{F}_\textsf{msb}$, $\mathcal{F}_\textsf{msbTOwrap}$, $\mathcal{F}_\textsf{AND}$, $\mathcal{F}_\textsf{Mux}$, $\mathcal{F}_{TR}$ has been proved in previous work~\cite{rathee2021sirnn}. Therefore, the security of the proposed $\mathcal{F}_{\textsf{exp}}$ follows in the $(\mathcal{F}_\textsf{CrossTerm}$, $\mathcal{F}_\textsf{msb}$, $\mathcal{F}_\textsf{msbTOwrap}$, $\mathcal{F}_\textsf{AND}$, $\mathcal{F}_\textsf{Mux}$, $\mathcal{F}_{TR})$-hybrid.

\subsection{Exponential Function with Negative Input only}\label{sec:ExponentialN}

\textbf{Motivation}. \textsf{Sigmoid} and \textsf{Tanh} are two common activation functions in neural networks. Both the \textsf{Sigmoid} and \textsf{Tanh} functions invoke the exponential function for computation. It can be observed that these functions can be transformed to piecewise functions by appropriate math variations. With this, we can only deal with one piece of the functions in 2PC scenario and the other pieces can be efficiently solved by the symmetry of the function. Here, we design a secure 2PC protocol for exponential function with negative input only, denoted as $\mathcal{F}_{\textsf{expn}}$. With the limited input, we can design a protocol with better efficiency than $\mathcal{F}_{\textsf{exp}}$.

\noindent \textbf{Design}. Give a negative float number $x_f \in \mathbb{R}$, we transform it to be a fixed-point number $x\in \mathbb{Z}_L$ with $L=2^l$ by $\textsf{Fix}$. Accordingly, $x$ can be reversed to be a float number $x'_f$ by $\textsf{rev}$ as in \ref{sec:Fixed-Point}. To construct the secure 2PC protocol, the fixed-point number $x$ is split into two additive secret shares $[{x}]_1$ and $[{x}]_2$ held by $P_1$ and $P_2$, respectively. We have $\textsf{msb}(x)$ always to be 1 when $x'_f$ is a negative number. For simplification, we write $\textsf{wrap}$ and $\textsf{msb}$ for $\textsf{wrap}(x)$ and $\textsf{msb}(x)$, respectively. Then, we list all possible cases of $x'_f$ and $e^{x'_f}$ with $\textsf{msb}= 1$ in Table~\ref{tab:sfpcase2} 
\begin{table}[!t]
\caption{The association from \textsf{wrap} to $x'_f$ and $e^{x'_f}$ with \textsf{msb} always to be 1.}
\centering
\setlength{\tabcolsep}{4.6mm}{
\begin{tabular}{|c|c|l|l|}
\hline
\textsf{wrap} & \textsf{msb} & \multicolumn{1}{c|}{$x'_f$} & \multicolumn{1}{c|}{$e^{x'_f}$}  \\ \hline
1 & 1 & ~~$\frac{[x]_1+[x]_2-2L}{2^s}$ & ~~$e^{\frac{[x]_1+[x]_2-2L}{2^s}}$  \\ \hline
0 & 1 & ~~$\frac{[x]_1+[x]_2-L}{2^s}$ & ~~$e^{\frac{[x]_1+[x]_2-L}{2^s}}$  \\ \hline
\end{tabular}
}
\label{tab:sfpcase2}
\end{table}
\noindent , which is derived from Table~\ref{tab:sfpcase}. According to Eq.~(\ref{equ:rev1}), we have
\begin{equation}
\label{equ:rev2}
\begin{aligned}
    x'_f    &=\frac{[x]_1+[x]_2-\textsf{wrap}\cdot L-\textsf{msb}\cdot L}{2^s}, \\
            &=\frac{[x]_1+[x]_2-L-\textsf{wrap}\cdot L}{2^s},
\end{aligned}
\end{equation}
with $\textsf{msb}=1 $. It can be further deduced that
\begin{equation}
\label{equ:Exponentn}
\begin{aligned}
    e^{x'_f}    &=e^{\frac{[x]_1+[x]_2-2L}{2^s}} \cdot \textsf{wrap}\\
                &+e^{\frac{[x]_1+[x]_2- L}{2^s}} \cdot (\textsf{wrap} \oplus 1),
\end{aligned}
\end{equation}
which, similar to Eq.~(\ref{equ:ExponentOpt}), can be optimized to be
\begin{equation}
\label{equ:ExponentnOpt}
\begin{aligned}
    e^{x'_f}    &=(e^{\frac{[x]_1+[x]_2-2L}{2^s}}-e^{\frac{[x]_1+[x]_2-L}{2^s}}) \cdot \textsf{wrap} \\
                &+ e^{\frac{[x]_1+[x]_2- L}{2^s}}.
\end{aligned}
\end{equation}
\noindent Our realization details of $\mathcal{F}_{\textsf{expn}}$ is shown in Algorithm~\ref{alg:ExponentialN}.

\begin{algorithm}[t]
\caption{Secure 2PC protocol for exponential function with negative input $\mathcal{F}_{\textsf{expn}}$} %
\label{alg:ExponentialN}
\begin{algorithmic}[1]
\REQUIRE
    For $i \in \{1,2\}$, $P_i$ holds $[{x}]_i$.\\
\ENSURE
    For $i \in \{1,2\}$, $P_i$ learns $[{y}]_i$.\\

    \STATE$P_i$ sets: \\
    $\quad a_i=\textsf{Fix}(e^{\frac{[x]_i-L/2}{2^s}},s')$,\\
    $\quad b_i=\textsf{Fix}(e^{\frac{[x]_i-L  }{2^s}},s')$, where $s' > s$ and will be discussed in Section~\ref{sec:ULPerrors}.
    \STATE $P_i$ invokes $\mathcal{F}_\textsf{CrossTerm}$ with inputs $a_i$, $b_i$ to learn $[a']_i$, $[b']_i$, with $a'=a_1 \times a_2$ and $b'=b_1 \times b_2$.
    \STATE $P_1$ sets $\left \langle\textsf{msb}\right \rangle_1$=0; $P_2$ sets: $\left \langle\textsf{msb}\right \rangle_2$=1.
    \STATE$P_i$ invokes $\mathcal{F}_\textsf{msbTOwrap}$ with inputs $\left \langle \textsf{msb} \right \rangle_i$ and $[{x}]_i$ to learn output $\left \langle\textsf{wrap}\right \rangle_i$.
    \STATE$P_i$ sets: $[{tmp1}]_i=[{b'}]_i - [{a'}]_i~mod~N$, where $N \ge L$ and will be discussed in Section~\ref{sec:implementation}.
    \STATE$P_i$ invokes $\mathcal{F}_\textsf{Mux}$ with inputs $\left \langle \textsf{wrap}\right \rangle_i$ and $[{tmp1}]_i$ to learn $[{tmp2}]_i$.
    \STATE$P_i$ sets: $[{rst}]_i =[{tmp2}]_i + [{a'}]_i~mod~N$.
    \STATE$P_i$ invokes $\mathcal{F}_\textsf{TR}$ with inputs $[{rst}]_i$ and common parameter $2s'-s$ to learn final output $[{y}]_i$.
\end{algorithmic}
\end{algorithm}

\noindent \textbf{Security analysis}. There are no interactions other than those generated by $\mathcal{F}_\textsf{CrossTerm}$, $\mathcal{F}_\textsf{msbTOwrap}$, $\mathcal{F}_\textsf{mux}$, and $\mathcal{F}_{TR}$. The $[{y}]_i$ is randomly uniform and the security of $\mathcal{F}_\textsf{CrossTerm}$, $\mathcal{F}_\textsf{msbTOwrap}$, $\mathcal{F}_\textsf{Mux}$, $\mathcal{F}_{TR}$ has been proved in previous work, therefore, the security of the $\mathcal{F}_{\textsf{expn}}$ follows in the $(\mathcal{F}_\textsf{CrossTerm}$, $\mathcal{F}_\textsf{msbTOwrap}$, $\mathcal{F}_\textsf{Mux}$, $\mathcal{F}_{TR})$-hybrid.

\subsection{Sigmoid Function}\label{sec:Sigmoid}
\noindent \textbf{Motivation}. $\textsf{sigmoid}$ is one of the most commonly used activation functions in neural networks, and is defined as
\begin{equation}\label{equ:sigmoid}
    \textsf{sigmoid}(x_f)=\frac{1}{1+e^{-x_f}}.
\end{equation}

 The computation of $\textsf{sigmoid}(x_f)$ can be naturally divided into the following three steps: first, calculate $e^{-x_f}$; then, $1+e^{-x_f}$; and finally, calculate the reciprocal of $1+e^{-x_f}$ to obtain the final result. In secure 2PC scenario, $e^{-x_f}$ can be computed by our proposed $\mathcal{F}_\textsf{exp}$ protocol. The addition operation $1+e^{-x_f}$ on secret shares can be done locally on two parties. Then, the calculation of reciprocal of $1+e^{-x_f}$ is the major challenge for the secure \textsf{sigmoid} protocol.

 The multiplication-based iterative algorithm is the usual method to compute reciprocal, which converges quadratically~\cite{goldschmidt1964applications,ito1997efficient}. However, the multiplication-based iterative algorithm requires an approximate estimate of result to initialize the iterative algorithm, so as to determine an $\alpha$ such that $2^{\alpha}\le v < 2^{\alpha+1}$, where $v$ is the number for which we need to calculate the reciprocal. It is a time-consuming operation.

\noindent \textbf{Design}. It can be found that we always have $2^0 \le 1+e^{-x_f} < 2^1$ with $x_f > 0$ and $2^0 \le 1+e^{x_f} < 2^1$ with $x_f < 0$. Accordingly, SIRNN \cite{rathee2021sirnn} transforms the sigmoid function into the following piecewise functions,
\begin{equation}
\label{equ:SigmoidPiece}
\begin{aligned}
\textsf{sigmoid}(x_f)&=\left \{
\begin{array}{ll}
    0.5,                                    & if\ x_f=0;\\ \\
    \frac{1}{1+e^{-x_f}},                   & if\ x_f>0;\\ \\
    e^{x_f}\cdot\frac{1}{1+e^{x_f}},      & if\ x_f<0.
\end{array} \right.
\end{aligned}
\end{equation}
It saves the cost incurred in computing $\alpha$, which is considerable expensive. Different from the previous work, we observe that Eq.~(\ref{equ:SigmoidPiece}) can be further optimized as
\begin{equation}
\label{equ:SigmoidOpt1}
\begin{aligned}
\textsf{sigmoid}(x_f)&=\left \{
\begin{array}{ll}
    0.5,                        & if\ x_f=0;\\ \\
    \frac{1}{1+e^{-x_f}},       & if\ x_f> 0;\\ \\
    1-\frac{1}{1+e^{x_f}},      & if\ x_f<0.
\end{array} \right.
\end{aligned}
\end{equation}

\noindent through the symmetry of the function. This optimization allows us to save one invocation to  $\mathcal{F}_\textsf{Mult}$  and $\mathcal{F}_\textsf{TR}$ compared to~\cite{rathee2021sirnn} during the implementation of the $\textsf{sigmoid}$ protocol.

 At first glance, Eq.~(\ref{equ:SigmoidOpt1}) can be further simplified to be

\begin{equation}
\label{equ:SigmoidOpt1-1}
\begin{aligned}
\textsf{sigmoid}(x_f)&=\left \{
\begin{array}{ll}
    \frac{1}{1+e^{-x_f}},       & if\ x_f \leq 0;\\ \\
    1-\frac{1}{1+e^{x_f}},      & if\ x_f<0,
\end{array} \right.
\end{aligned}
\end{equation}

 \noindent which can further save one invocation to  $\mathcal{F}_\textsf{msb}$  and $\mathcal{F}_\textsf{B2A}$. But please note that we have $1 + e^{x_f} =2 $ when $x_f = 0$, which does not satisfy the requirement $(1 + e^{x_f}<2)$ in multiplication-based iterative algorithm. Different from the previous work, we introduce a trivial error to solve this problem. In our protocol, a floating-point number $x_f$ will be transformed to be its fixed-point representation $x$. Then, let $x \leftarrow x + 1$ when $\textsf{msb} = 0$, which means we add a tiny error $\varepsilon=2^{-s}$ to all positive number $x'_f$ which denotes the float number transformed back from the fixed-point number $x$. Then, the function Eq.~(\ref{equ:SigmoidOpt1-1}) can be transformed into  the following piecewise functions,
\begin{equation}
\label{equ:SigmoidOpt2}
\begin{aligned}
\textsf{sigmoid}(x'_f)&=\left \{
\begin{array}{ll}
    \frac{1}{1+e^{-x'_f-\varepsilon}},       & if\ x'_f\ge 0\\ \\
    1-\frac{1}{1+e^{x'_f}},     & if\ x'_f<0.
\end{array} \right.
\end{aligned}
\end{equation}
\noindent which can further save one invocation to  $\mathcal{F}_\textsf{msb}$  and $\mathcal{F}_\textsf{B2A}$ compared to~\cite{rathee2021sirnn} during the implementation of the $\textsf{sigmoid}$ protocol. For ease of expression, we define $\textsf{negx}$ as
\[\textsf{negx}=
\begin{cases}
    -(x+1), &if\ \textsf{msb}=0,\\
    x, &if\ \textsf{msb}=1.
\end{cases}\]
It is not hard to deduce that
\begin{equation}
\label{equ:negx}
\begin{aligned}
\textsf{rev(negx)} =
    \begin{cases}
    -x'_f-\varepsilon ,&if\ \textsf{msb}=0, \\
    x'_f,&if\ \textsf{msb}=1,
    \end{cases}
\end{aligned}
\end{equation}
and $\textsf{rev(negx)}$ is always negative. According to Eq.~(\ref{equ:SigmoidOpt2}) and Eq.~(\ref{equ:negx}), we have
\begin{equation}
\nonumber
\begin{aligned}
\textsf{sigmoid}(x'_f)=
    \begin{cases}
    \frac{1}{1+e^{\textsf{rev(negx)}}},&if\ \textsf{msb}=0, \\ \\
    1-\frac{1}{1+e^{\textsf{rev(negx)}}},&if\ \textsf{msb}=1.
    \end{cases}
\end{aligned}
\end{equation}
Finally, our implementation
details of $\mathcal{F}_{\textsf{sigmoid}}$ is presented in Algorithm~\ref{alg:Sigmoid}.

\begin{algorithm}[t]
\caption{Secure 2PC sigmoid protocol $\mathcal{F}_{\textsf{sigmoid}}$}
\label{alg:Sigmoid}
\begin{algorithmic}[1]
\REQUIRE
    For $i \in \{1,2\}$, $P_i$ holds $[{x}]_i$.\\
\ENSURE
    For $i \in \{1,2\}$, $P_i$ learns $[{y}]_i$.

    \STATE $P_1$ sets:
    $\quad{[2x+1]}_1 = {2\cdot[x]}_1 + 1$;\\
    $P_2$ sets:
    $\quad{[2x+1]}_2 = {2\cdot[x]}_2$.
    \STATE $P_i$ invokes $\mathcal{F}_\textsf{msb}$ with input $[{x}]_i$ to learn output $\left \langle \textsf{msb}\right \rangle_i$.
    \STATE $P_i$ invokes $\mathcal{F}_\textsf{Mux}$ with inputs $\left \langle \textsf{msb}\right \rangle_i$ and ${[2x+1]}_i$ to learn $[(2x+1)\cdot \textsf{msb}]_i$.
    \STATE $P_1$ sets: ${[\textsf{negx}]}_1=[(2x+1) \cdot \textsf{msb}]_1-[x]_1-1$.\\
    $P_2$ sets: ${[\textsf{negx}]}_2=[(2x+1) \cdot \textsf{msb}]_2-[x]_2$.
    \STATE $P_i$ invokes $\mathcal{F}_\textsf{expn}$ with input ${[\textsf{negx}]}_i$ to learn $[\textsf{Fix}(e^{\textsf{rev}(\textsf{negx})},s)]_i$.
    \STATE $P_1$ sets:  $~{[\textsf{Fix}({1+e^{\textsf{rev(negx)}}},s)]_1}$\\$\quad\quad\quad~=\textsf{Fix}(1,s)+{[\textsf{Fix}(e^{\textsf{rev}(\textsf{negx})},s)]_1}$; \\ $P_2$ sets: $~{[\textsf{Fix}({1+e^{\textsf{rev(negx)}}},s)]_2}$\\$\quad\quad\quad~={[\textsf{Fix}(e^{\textsf{rev}(\textsf{negx})},s)]_2}$.
    \STATE $P_i$ invokes $\mathcal{F}_\textsf{Rec}$ with input $[\textsf{Fix}({1+e^{\textsf{rev(negx)}}},s)]_i$ to learn output $[\textsf{Fix}(\frac{1}{1+e^{\textsf{rev(negx)}}},s)]_i$.
    \STATE $P_1$ sets: $~{[\textsf{Fix}((1-2\cdot\frac{1}{1+e^{\textsf{rev(negx)}}},s)]_1}$\\$\quad\quad\quad~=\textsf{Fix}(1,s)-2\cdot{[\textsf{Fix}(\frac{1}{1+e^{\textsf{rev(negx)}}},s)]_1}$; \\
    $P_2$ sets:$~{[\textsf{Fix}((1-2\cdot\frac{1}{1+e^{\textsf{rev(negx)}}},s)]_2}$\\$\quad\quad\quad~=-2\cdot{[\textsf{Fix}(\frac{1}{1+e^{\textsf{rev(negx)}}},s)]_2}$.

    \STATE$P_i$ invokes $\mathcal{F}_\textsf{Mux}$ with inputs ${[\textsf{Fix}(1-2\cdot\frac{1}{1+e^{\textsf{rev(negx)}}},s)]_i}$ and $\left \langle \textsf{msb}\right \rangle_i$ to learn ${[\textsf{Fix}((1-2\cdot\frac{1}{1+e^{\textsf{rev(negx)}}})\cdot \textsf{msb},s)]_i}$.
    \STATE $P_i$ sets:
    $ [y]_i=[\textsf{Fix}(\frac{1}{1+e^{\textsf{rev(negx)}}},s)]_i$\\
    $\quad\quad\quad\quad~~ +{[{\textsf{Fix}((1-2\cdot\frac{1}{1+e^{\textsf{rev(negx)}}})\cdot \textsf{msb},s)}]_i}$.
\end{algorithmic}
\end{algorithm}

\noindent \textbf{Security analysis}. There are no interactions other than those generated by $\mathcal{F}_\textsf{msb}$, $\mathcal{F}_\textsf{mux}$, $\mathcal{F}_\textsf{expn}$, and $\mathcal{F}_{Rec}$. The $[{y}]_i$ is randomly uniform and the security of $\mathcal{F}_\textsf{msb}$, $\mathcal{F}_\textsf{mux}$, $\mathcal{F}_{Rec}$ has been proved in previous work. The security of $\mathcal{F}_\textsf{expn}$ has been proved in Section \ref{sec:ExponentialN}. Therefore, the security of the $\mathcal{F}_{\textsf{sigmoid}}$ follows in the $(\mathcal{F}_\textsf{msb}$, $\mathcal{F}_\textsf{mux}$, $\mathcal{F}_\textsf{expn}$, $\mathcal{F}_{Rec})$-hybrid.

\subsection{Tanh Function}\label{sec:Tanh}
\noindent \textbf{Motivation}. $\textsf{Tanh}$ is another most commonly used activation functions in neural networks, and is defined as
\[\textsf{Tanh}(x_f)=\frac{1-e^{-2x_f}}{1+e^{-2x_f}}.\]
In secure 2PC scenario, the intention of $P_i$ is to obtain the secret result of $\textsf{Fix}(\textsf{Tanh}(x'_f), s)$ without revealing any information about its $[{x}]_i$.

\noindent \textbf{Design}. The \textsf{Tanh} and \textsf{sigmoid} functions are closely related~\cite{goodfellow2016deep}: $\textsf{Tanh}(x_f) = 2\cdot\textsf{sigmoid}(2x_f)-1$. Thus, the implementation of $\mathcal{F}_\textsf{Tanh}$ can be realized by $\mathcal{F}_\textsf{sigmoid}$. The implementation
details of $\mathcal{F}_\textsf{Tanh}$ is presented in Algorithm~\ref{alg:Tanh}.

\begin{algorithm}[t]
\caption{Secure 2PC Tanh protocol $\mathcal{F}_{\textsf{Tanh}}$}
\label{alg:Tanh}
\begin{algorithmic}[1]
\REQUIRE
    For $i \in \{1,2\}$, $P_i$ holds $[{x}]_i$.\\
\ENSURE
    For $i \in \{1,2\}$, $P_i$ learns $[{y}]_i$.

    \STATE $P_1$ sets:
    ${[2x]}_1 = {2\cdot[x]}_1$;\\
    $P_1$ sets:
    ${[2x]}_2 = {2\cdot[x]}_2$.
    \STATE $P_i$ invokes $\mathcal{F}_\textsf{sigmoid}$ with input $[{2x}]_i$ to learn output $[\textsf{Fix}(\textsf{sigmoid}(\textsf{rev}(2x)),s)]_i$.
    \STATE $P_1$ sets: ${[y]_1}=2\cdot{[\textsf{Fix}(\textsf{sigmoid}(\textsf{rev}(2x)),s)]_1}-\textsf{Fix}(1,s)$; \\
    $P_2$ sets: ${[y]_2}=2\cdot{[\textsf{Fix}(\textsf{sigmoid}(\textsf{rev}(2x)),s)]_2}$.
\end{algorithmic}
\end{algorithm}

\noindent \textbf{Security analysis}. There are no interactions other than those generated by $\mathcal{F}_\textsf{sigmoid}$ and the $[y]_i$ is randomly uniform, so that the $\mathcal{F}_\textsf{Tanh}$ is as safe as $\mathcal{F}_\textsf{sigmoid}$. The security of $\mathcal{F}_\textsf{sigmoid}$ has been proved in Section~\ref{sec:Sigmoid}.

\section{Implementation}\label{sec:implementation}
 Our implementation is written in C++ and uses the fundamental 2PC protocols from SIRNN~\cite{rathee2021sirnn}, which are available in the EzPC library~\cite{EzPC}. Similar to SIRNN, our proposed protocols also accept non-uniform bit-width inputs and outputs. We provide a complete implementation and test code for the proposed 2PC protocol to confirm the correctness and feasibility of our approach, and the code will be open sourced on GitHub\footnote{https://github.com/Samsara430381/2PC.git.}.

 We set the same precision parameter $s$ for both input and output in our implementation, but in some steps of Algorithm~\ref{alg:Exponential} and~\ref{alg:ExponentialN}, we have to use a larger precision parameter $s'(>s)$ to increase the precision as the exponential function can magnify the error dramatically. Additionally, the $N$ in Algorithm~\ref{alg:Exponential} and~\ref{alg:ExponentialN} defines the size of ring $\mathbb{Z}_N$, which can be set to $2^{64}$ as a maximum. And we require that the real floating-point result not exceed $N$ after scaling $2^{2s'}\times$. Accordingly, the selection of $s'$ is limited by both $N$ and the range of the result. As a result, we can leave 4 bits to express the integer parts. Also, in the implementation of $\mathcal{F}_\textsf{sigmoid}$ and $\mathcal{F}_\textsf{Tanh}$ as in Algorithm~\ref{alg:Sigmoid} and~\ref{alg:Tanh}, it is necessary to use SIRNN's $\mathcal{F}_\textsf{expn}$ protocol for inputs with integer parts exceeding 4 bits.

For 2PC RNN inference, we followed SIRNN's construction by just replacing the protocols for $\mathcal{F}_\textsf{sigmoid}$ and $\mathcal{F}_\textsf{Tanh}$ with ours in EzPC library. In addition, we used SeeDot~\cite{gopinath2019compiling} to convert the floating-point code of the FastGRNN~\cite{kusupati2018fastgrnn} network architecture into the EzPC code. The generated Ezpc code is then automatically compiled into a secure and efficient computation protocol for the same code by EzPC framework.

\section{Evaluation}

\begin{table*}[!t]
\caption{The Max UPL Errors of the protocols with various $s'$.}
\label{tab:ULPsOnS}
\centering
\setlength{\tabcolsep}{1.49mm}{
\begin{tabular}{|c|cccccccccccccccc|}
\hline
\multirow{2}{*}{\textbf{Functions}} & \multicolumn{16}{c|}{\textbf{Max ULP Errors on a variety of $s'$}}                                                                                                                                                                                                                                                                                                                    \\ \cline{2-17}
& \multicolumn{1}{c|}{$s+0$}
& \multicolumn{1}{c|}{$s+1$}
& \multicolumn{1}{c|}{$s+2$}
& \multicolumn{1}{c|}{$s+3$}
& \multicolumn{1}{c|}{$s+4$}
& \multicolumn{1}{c|}{$s+5$}
& \multicolumn{1}{c|}{$s+6$}
& \multicolumn{1}{c|}{$s+7$}
& \multicolumn{1}{c|}{$s+8$}
& \multicolumn{1}{c|}{$s+9$}
& \multicolumn{1}{c|}{$s+10$}
& \multicolumn{1}{c|}{$s+11$}
& \multicolumn{1}{c|}{$s+12$}
& \multicolumn{1}{c|}{$s+13$}
& \multicolumn{1}{c|}{$s+14$}
& $s+15$
\\ \hline
\textsf{$\mathcal{F}_{\textsf{exp}}$}
& \multicolumn{1}{c|}{2977}
& \multicolumn{1}{c|}{1484}
& \multicolumn{1}{c|}{745}
& \multicolumn{1}{c|}{372}
& \multicolumn{1}{c|}{186}
& \multicolumn{1}{c|}{93}
& \multicolumn{1}{c|}{46}
& \multicolumn{1}{c|}{24}
& \multicolumn{1}{c|}{12}
& \multicolumn{1}{c|}{6}
& \multicolumn{1}{c|}{-}
& \multicolumn{1}{c|}{-}
& \multicolumn{1}{c|}{-}
& \multicolumn{1}{c|}{-}
& \multicolumn{1}{c|}{-}
& -
\\ \hline
\textsf{$\mathcal{F}_{\textsf{expn}}$}
& \multicolumn{1}{c|}{2935}
& \multicolumn{1}{c|}{1421}
& \multicolumn{1}{c|}{722}
& \multicolumn{1}{c|}{349}
& \multicolumn{1}{c|}{180}
& \multicolumn{1}{c|}{88}
& \multicolumn{1}{c|}{44}
& \multicolumn{1}{c|}{22}
& \multicolumn{1}{c|}{11}
& \multicolumn{1}{c|}{6}
& \multicolumn{1}{c|}{3}
& \multicolumn{1}{c|}{2}
& \multicolumn{1}{c|}{1}
& \multicolumn{1}{c|}{1}
& \multicolumn{1}{c|}{1}
& 1
\\ \hline
\textsf{$\mathcal{F}_{\textsf{sigmoid}}$}
& \multicolumn{1}{c|}{749}
& \multicolumn{1}{c|}{369}
& \multicolumn{1}{c|}{185}
& \multicolumn{1}{c|}{93}
& \multicolumn{1}{c|}{46}
& \multicolumn{1}{c|}{23}
& \multicolumn{1}{c|}{12}
& \multicolumn{1}{c|}{6}
& \multicolumn{1}{c|}{4}
& \multicolumn{1}{c|}{3}
& \multicolumn{1}{c|}{3}
& \multicolumn{1}{c|}{3}
& \multicolumn{1}{c|}{3}
& \multicolumn{1}{c|}{3}
& \multicolumn{1}{c|}{3}
& 3
\\ \hline
\textsf{$\mathcal{F}_{\textsf{Tanh}}$}
& \multicolumn{1}{c|}{3048}
& \multicolumn{1}{c|}{1492}
& \multicolumn{1}{c|}{735}
& \multicolumn{1}{c|}{362}
& \multicolumn{1}{c|}{182}
& \multicolumn{1}{c|}{92}
& \multicolumn{1}{c|}{47}
& \multicolumn{1}{c|}{24}
& \multicolumn{1}{c|}{12}
& \multicolumn{1}{c|}{7}
& \multicolumn{1}{c|}{5}
& \multicolumn{1}{c|}{4}
& \multicolumn{1}{c|}{4}
& \multicolumn{1}{c|}{4}
& \multicolumn{1}{c|}{4}
& \multicolumn{1}{c|}{4}
\\ \hline
\end{tabular}
}
\end{table*}

 This section tests the accuracy and efficiency of the proposed protocols. And we have successfully implemented the state-of-the-art 2PC FastGRNN~\cite{kusupati2018fastgrnn} inference on Google-30 dataset with the proposed protocols.

 \textbf{Experiment Setup}. The experiments are conducted on a computer with an Intel i9-12900KF CPU and 128GB RAM. We created two docker containers with ubuntu 20.04 image, which represent the two participants $P_1$ and $P_2$. To simulate LAN environment and WAN environment respectively, we use traffic control (TC) command to set port speed limit. The bandwidth between $P_1$ and $P_2$ is about 358 MBps for LAN and 45 MBps for WAN, respectively. The round-trip time is about 0.3~ms for LAN and 40~ms for WAN, respectively. Each experiment was repeated more than 5 times, and we calculated the average of the maximum and minimum latency as the latency result.

  \textbf{Benchmarks}. We compared our scheme in terms of precision and efficiency with SIRNN~\cite{rathee2021sirnn}, which is the most advanced 2PC system running in semi honest settings. For precision, we use the ULP errors as benchmarks; and for efficiency, the running time is the final index which is determined by the communication loads and number of communication rounds. Generally, given a task one can reduce the running time by packing the communication loads together. With a large \textit{batch size}, the communication time will be determined by communication load; while with a small \emph{batch size}, the communication time will be consumed the number of communication rounds.
  Finally, we applied our \textsf{sigmoid} and \textsf{Tanh} protocols to an end-to-end secure RNN inference, and also compare the performance with SIRNN~\cite{rathee2021sirnn}.

\subsection{ULP Errors}\label{sec:ULPerrors}

 In this section, we evaluate the precision of the proposed protocols with Max ULP errors, which is an established precision metric in many previous works. In our evaluations, we represent the input by a bit length of $ s+4$. Here $s$ is the magnifying scale as in Eq.~\ref{equ:Fix}, which defines the precision of input reserved. Then, 4 bits is leaved to represent the integer part of input, which determines the range of input in (-8, 8). We set the same $s$ for output to get the same precision. For efficiency, we set the bit-width of output to the minimum length which is able to correctly represent the output. Specifically, the bit-width of output ($bw_o$) in $\mathcal{F}_{\textsf{exp}}$ is set to $s+13$ and the bit-width of output in $\mathcal{F}_{\textsf{expn}}$, $\mathcal{F}_{\textsf{sigmoid}}$, and $\mathcal{F}_{\textsf{Tanh}}$ are all set to $s+2$.

 As mentioned in Section~\ref{sec:implementation}, in some steps of Algorithm~\ref{alg:Exponential} and~\ref{alg:ExponentialN}, we have to use a larger precision parameter $s'(>s)$ to increase the precision as the exponential function can magnify the error dramatically. However, our implementation does not allow $bw_o-s+2s'$ to exceed 64 where $bw_o$ denotes the output length. Accordingly, the we can only set $s'=s+9$ as a maximum for $\mathcal{F}_{\textsf{exp}}$ and obtain the Max ULP Error of 6. For $\mathcal{F}_{\textsf{expn}}$, $\mathcal{F}_{\textsf{sigmoid}}$, and $\mathcal{F}_{\textsf{Tanh}}$, we can set $s'=s+15$ and obtain the Max ULP Error of 1, 3, and 4, respectively.

 To ensure the accuracy of the results, we set $s'$ to $s+9$, $s+12$, $s+9$, and $s+11$ for $\mathcal{F}_{\textsf{exp}}$, $\mathcal{F}_{\textsf{expn}}$, $\mathcal{F}_{\textsf{sigmoid}}$, and $\mathcal{F}_{\textsf{Tanh}}$,  respectively, in the following evaluation. With this setting, we compare the precision of our protocols with some previous ones. As shown in Table~\ref{tab:ULPs}, we achieve a state-of-the-art precision on $\mathcal{F}_{\textsf{exp}}$, $\mathcal{F}_{\textsf{expn}}$, $\mathcal{F}_{\textsf{sigmoid}}$, and $\mathcal{F}_{\textsf{Tanh}}$.

\begin{table}[!t]
\caption{The Max UPL Errors of the protocols compared with prior works.}
\label{tab:ULPs}
\centering
\setlength{\tabcolsep}{3.4mm}{
\begin{tabular}{|c|cccc|}
\hline
\multicolumn{1}{|c|}{\multirow{2}{*}{\textbf{Technique}}}
& \multicolumn{4}{c|}{\textbf{Max Errors} \textsf{(in ULP)}}
\\ \cline{2-5}
\multicolumn{1}{|c|}{}
& \multicolumn{1}{|c|}{$\mathcal{F}_{\textsf{exp}}$}
& \multicolumn{1}{c|}{$\mathcal{F}_{\textsf{expn}}$}
& \multicolumn{1}{c|}{$\mathcal{F}_{\textsf{sigmoid}}$}
& \multicolumn{1}{c|}{$\mathcal{F}_{\textsf{Tanh}}$}
\\ \hline
\textsf{SecureML}~\cite{mohassel2017secureml}
& \multicolumn{1}{c|}{-}
& \multicolumn{1}{c|}{-}
& \multicolumn{1}{c|}{1547}
& - \\ \hline
\makecell[c]{\textsf{MiniONN}~\cite{liu2017oblivious}\\12-piece}
& \multicolumn{1}{c|}{-}
& \multicolumn{1}{c|}{-}
& \multicolumn{1}{c|}{104}
& - \\ \hline
\makecell[c]{\textsf{MiniONN}~\cite{liu2017oblivious}\\48-piece}
& \multicolumn{1}{c|}{-}
& \multicolumn{1}{c|}{-}
& \multicolumn{1}{c|}{4}
& - \\ \hline
\textsf{SIRNN~\cite{rathee2021sirnn}}
& \multicolumn{1}{c|}{-}
& \multicolumn{1}{c|}{3}
& \multicolumn{1}{c|}{3}
& 4 \\ \hline
\textsf{MP-SPDZ~\cite{keller2020mp}}
& \multicolumn{1}{c|}{10557}
& \multicolumn{1}{c|}{2}
& \multicolumn{1}{c|}{2}
& - \\ \hline
\textsf{Ours}
& \multicolumn{1}{c|}{6}
& \multicolumn{1}{c|}{1}
& \multicolumn{1}{c|}{3}
& 4 \\ \hline
\end{tabular}
}
\end{table}

\subsection{Number of Building Blocks}\label{sec:Buildingblocks}
\begin{table*}[!t]
\caption{The required building blocks for the protocols. $l$ denotes the bit length of the input digit and $d$ is set as 8 by default.}
\label{tab:BuildingBlocks}
\centering
\setlength{\tabcolsep}{10.2mm}{
\begin{tabular}{|c|c|c|}
\hline
\multicolumn{1}{|l|}{\textbf{Protocol}} & \textbf{Technique} & \multicolumn{1}{c|}{\textbf{Building blocks}}
\\ \hline
\multirow{2}{*}{$\mathcal{F}_{\textsf{exp}}$}
& -
& -
\\ \cline{2-3}
& \textsf{Ours}
& $\mathcal{F}_{\textsf{CrossTerm}} \times 3$,~$\mathcal{F}_{\textsf{msb}} \times 1$,~ $\mathcal{F}_{\textsf{msbTOwrap}} \times 1$,~$\mathcal{F}_{\textsf{AND}} \times 1$,~ $\mathcal{F}_{\textsf{Mux}} \times 2$,~$\mathcal{F}_{\textsf{TR}} \times 1$
\\ \hline
\multirow{2}{*}{$\mathcal{F}_{\textsf{expn}}$}
& \textsf{SIRNN~\cite{rathee2021sirnn}}
& $\mathcal{F}_{\textsf{DigDec}} \times 1$,~$\mathcal{F}_{\textsf{LUT}} \times \lceil{l/d}\rceil$,~$\mathcal{F}_{\textsf{Mult}} \times (\lceil{l/d}\rceil-1)$,~$\mathcal{F}_{\textsf{TR}} \times (\lceil{l/d}\rceil-1)$
\\ \cline{2-3}
& \textsf{Ours}
&  $\mathcal{F}_{\textsf{CrossTerm}} \times 2$,~$\mathcal{F}_{\textsf{msbTOwrap}} \times 1$,~ $\mathcal{F}_{\textsf{Mux}} \times 1$,~$\mathcal{F}_{\textsf{TR}} \times 1$
\\ \hline
\multirow{2}{*}{$\mathcal{F}_{\textsf{sigmoid}}$}
& \textsf{SIRNN~\cite{rathee2021sirnn}}
& $\mathcal{F}_{\textsf{msb}} \times 2$,~$\mathcal{F}_{\textsf{Mux}} \times 2$,~ $\mathcal{F}_{\textsf{Rec}} \times 1$,~$\mathcal{F}_{\textsf{expn}} \times 1$,~ $\mathcal{F}_{\textsf{B2A}} \times 1$,~$\mathcal{F}_{\textsf{Mult}} \times 1$,~ $\mathcal{F}_{\textsf{TR}} \times 1$
\\ \cline{2-3}
& \textsf{Ours}
& $\mathcal{F}_{\textsf{msb}} \times 1$,~$\mathcal{F}_{\textsf{Mux}} \times 2$,~ $\mathcal{F}_{\textsf{Rec}} \times 1$,~$\mathcal{F}_{\textsf{expn}} \times 1$
\\ \hline
\multirow{2}{*}{$\mathcal{F}_{\textsf{Tanh}}$}
& \textsf{SIRNN~\cite{rathee2021sirnn}}
& $\mathcal{F}_{\textsf{msb}} \times 2$,~$\mathcal{F}_{\textsf{Mux}} \times 2$,~ $\mathcal{F}_{\textsf{Rec}} \times 1$,~$\mathcal{F}_{\textsf{expn}} \times 1$,~ $\mathcal{F}_{\textsf{B2A}} \times 1$,~$\mathcal{F}_{\textsf{Mult}} \times 1$,~ $\mathcal{F}_{\textsf{TR}} \times 1$
\\ \cline{2-3}
& \textsf{Ours}
& $\mathcal{F}_{\textsf{msb}} \times 1$,~$\mathcal{F}_{\textsf{Mux}} \times 2$,~ $\mathcal{F}_{\textsf{Rec}} \times 1$,~$\mathcal{F}_{\textsf{expn}} \times 1$
\\ \hline
\end{tabular}
}
\end{table*}

 In this section, we compare the number of building blocks invoked in protocols as shown in Table~\ref{tab:BuildingBlocks}. For $\mathcal{F}_{\textsf{expn}}$, we take a distinct technical route, and therefore have significant differences on building blocks. SIRNN used $\mathcal{F}_{\textsf{DigDec}}$ to decompose an $l$-bit number into $c$ sub-strings of length $d$, where the $d$ is set as 8 by default. It can be realized as in~\cite{rathee2021sirnn} with $(l/d- 1)(\lambda(d + 2) + 15d + 20)$ bits of total communication. SIRNN designed $\mathcal{F}_{\textsf{LUT}}$ to take a $d$-bits number $x$ as input and output a $n$-bits result $y$ such that $y=T(x)$. It can be realized using a single call to $\binom{2^d}{1}$-$\textsf{OT}_n$ with communication $2\lambda + 2^dn$ bits~\cite{dessouky2017pushing}. To sum up, our protocol needs fewer building blocks both in types and numbers, and achieves less communication traffic theoretically.

\subsection{Communication Rounds}\label{sec:Commrounds}

\begin{table}[!t]
\caption{The communication rounds of the protocols with $s$ = 12 and 16.}
\label{tab:Rounds}
\centering
\setlength{\tabcolsep}{2.5mm}{
\begin{tabular}{|c|c|cccc|}
\hline
\multirow{2}{*}{\textbf{Technique}}
& \multirow{2}{*}{\textbf{Scale}}
& \multicolumn{4}{c|}{\textbf{Rounds}}                                                                                                         \\ \cline{3-6}
&
& \multicolumn{1}{c|}{$\mathcal{F}_{\textsf{exp}}$}
& \multicolumn{1}{c|}{$\mathcal{F}_{\textsf{expn}}$}
& \multicolumn{1}{c|}{$\mathcal{F}_{\textsf{sigmoid}}$}
& $\mathcal{F}_{\textsf{Tanh}}$
\\ \hline
\multirow{2}{*}{\textsf{MP-SPDZ~\cite{keller2020mp}}}
& 12
& \multicolumn{1}{c|}{24}
& \multicolumn{1}{c|}{24}
& \multicolumn{1}{c|}{56}
& -
\\ \cline{2-6}
& 16
& \multicolumn{1}{c|}{34}
& \multicolumn{1}{c|}{34}
& \multicolumn{1}{c|}{62}
& -
\\ \hline
\multirow{2}{*}{\textsf{SIRNN~\cite{rathee2021sirnn}}}
& 12
& \multicolumn{1}{c|}{-}
& \multicolumn{1}{c|}{26}
& \multicolumn{1}{c|}{97}
& 91
\\ \cline{2-6}
& 16
& \multicolumn{1}{c|}{-}
& \multicolumn{1}{c|}{46}
& \multicolumn{1}{c|}{117}
& 109
\\ \hline
\multirow{2}{*}{\textsf{Ours}}
& 12
& \multicolumn{1}{c|}{\multirow{2}{*}{23}}
& \multicolumn{1}{c|}{\multirow{2}{*}{12}}
& \multicolumn{1}{c|}{\multirow{2}{*}{61}}
& \multirow{2}{*}{61} \\ \cline{2-2}
& 16
& \multicolumn{1}{c|}{}
& \multicolumn{1}{c|}{}
& \multicolumn{1}{c|}{}
&
\\ \hline
\end{tabular}
}
\end{table}

 In this section, we test the required communication rounds with prior works with $s$ = 12 and 16, respectively. Theoretically, our protocols only need a constant rounds of communication which is fewer than that in SIRNN~\cite{rathee2021sirnn} as shown in Table~\ref{tab:Rounds}. Compared with MP-SPDZ~\cite{keller2020mp}, we have fewer communication rounds in $\mathcal{F}_{\textsf{expn}}$ and similar communication rounds in $\mathcal{F}_{\textsf{exp}}$ and $\mathcal{F}_{\textsf{sigmoid}}$.

\subsection{Communication Traffic}\label{sec:Commloads}
 This section tests and compares the communication traffic. In real application, we can reduce communication burden by packing the communication loads together. Here, we pack the communication of 16384 instances of protocol together. And the communication traffic is taken as the averaged of these instances.

\begin{figure}
    \begin{center}
    \centering 
    \includegraphics[width=1\linewidth]{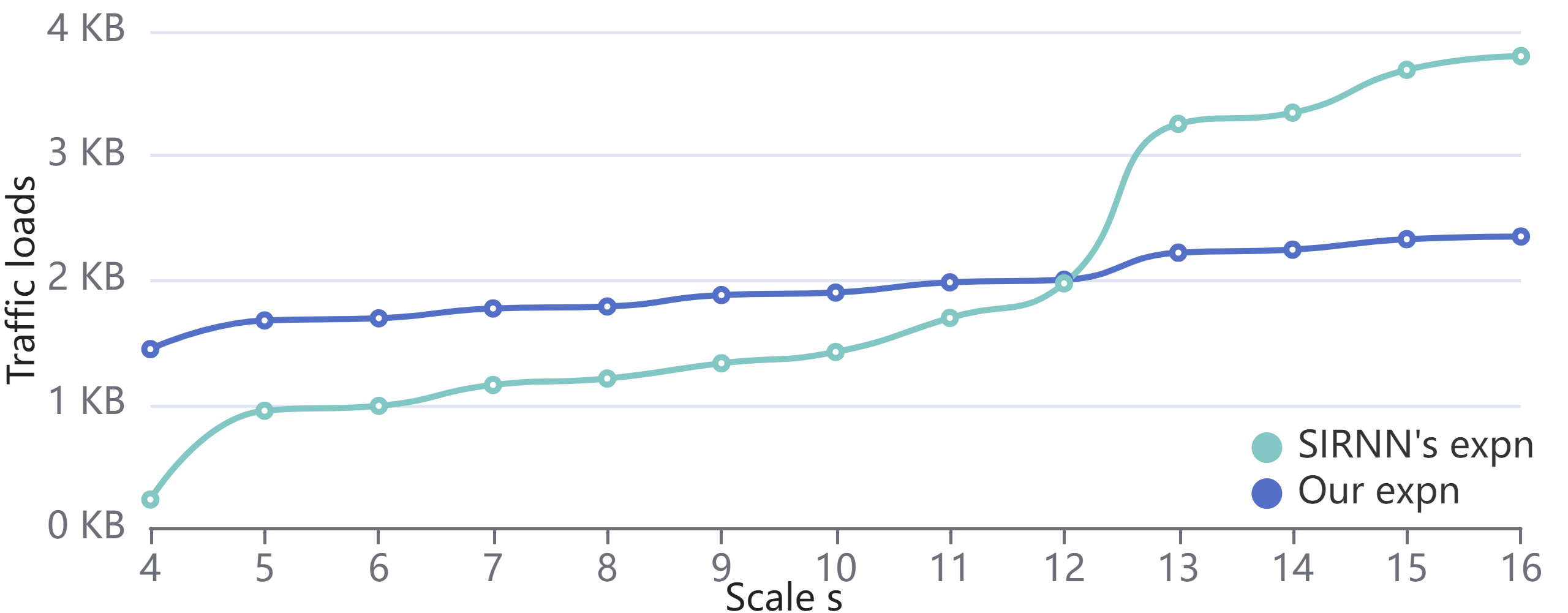}
    \end{center}
    \caption{\label{fig:CommofExpn} The communication traffic of $\mathcal{F}_\textsf{expn}$ with different scale $s$.}
    \end{figure}

\begin{figure}
    \begin{center}
    \centering 
    \includegraphics[width=1\linewidth]{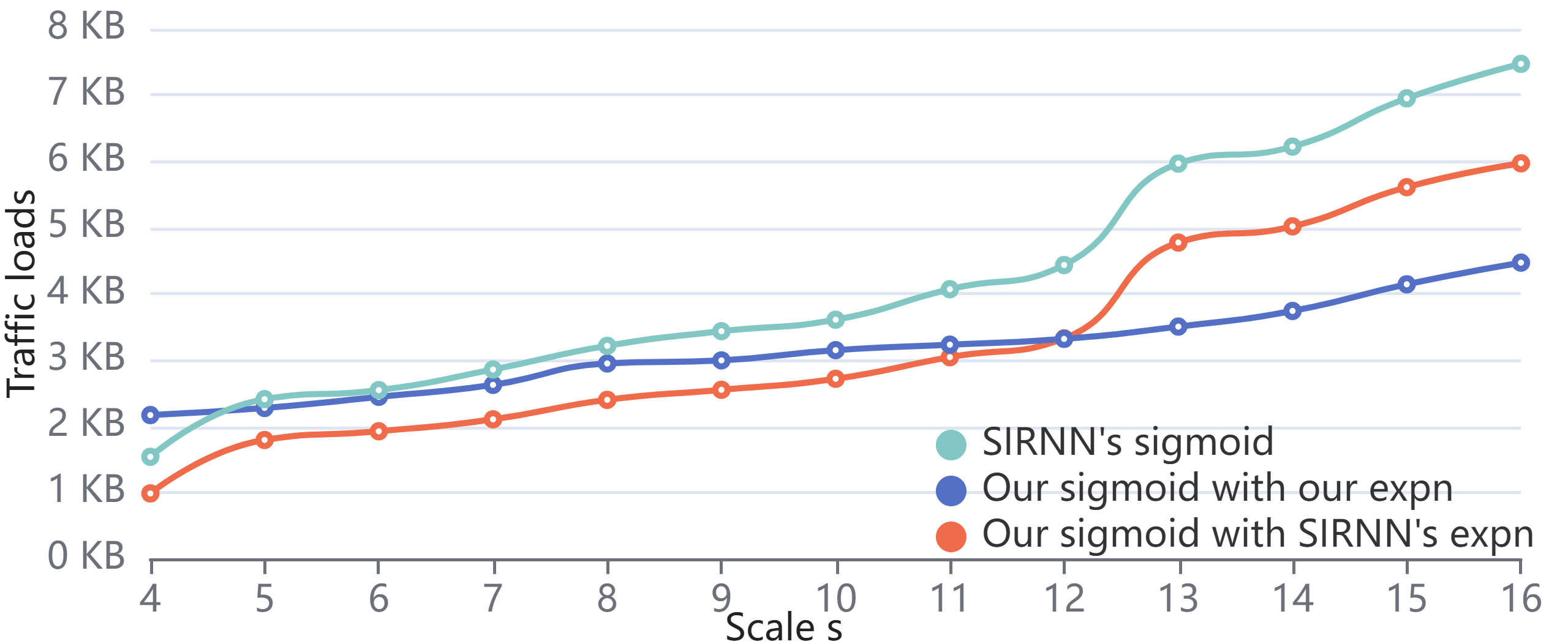}
    \end{center}
    \caption{\label{fig:SigmoidComm} The communication traffic of $\mathcal{F}_\textsf{sigmoid}$ with different scale $s$.}
    \end{figure}

\begin{figure}
    \begin{center}
    \centering 
    \includegraphics[width=1\linewidth]{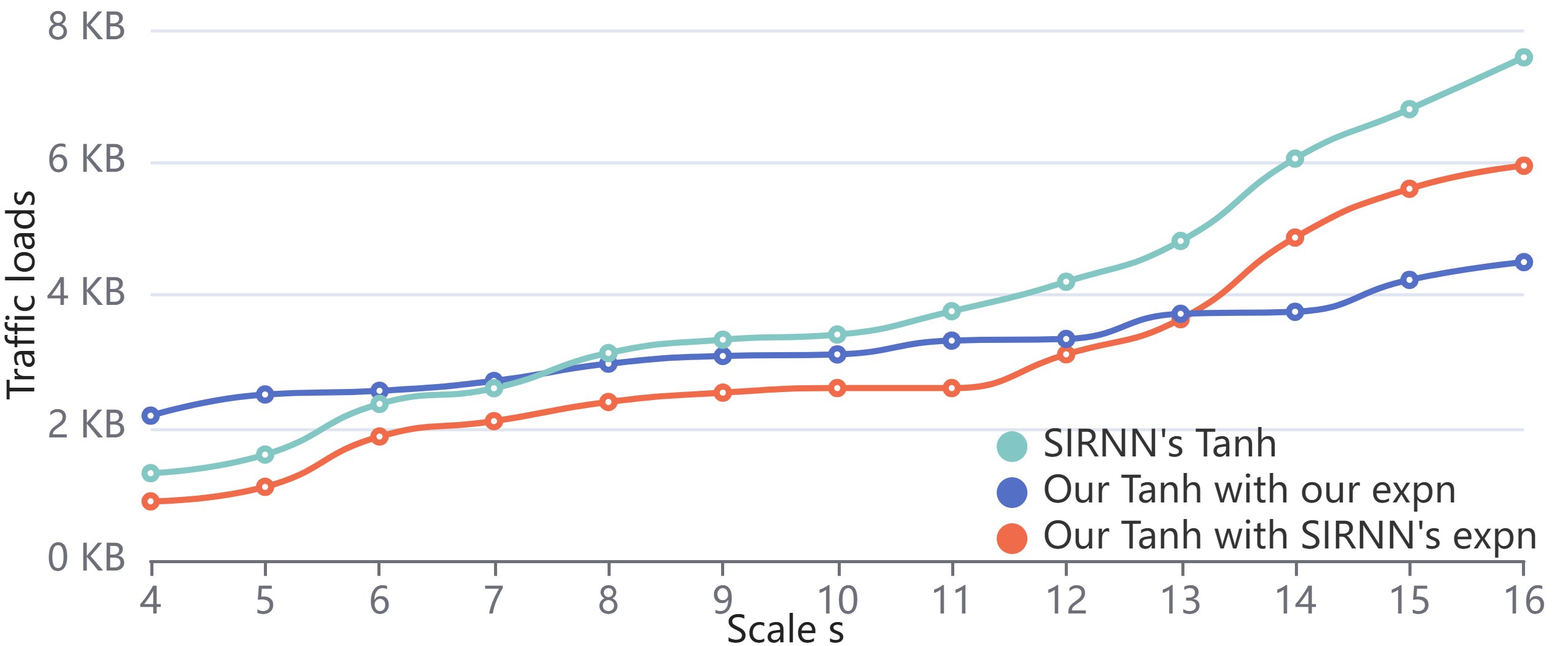}
    \end{center}
    \caption{\label{fig:TanhComm} The communication traffic of $\mathcal{F}_\textsf{Tanh}$ with different scale $s$.}
    \end{figure}

 As shown in Fig.~\ref{fig:CommofExpn}, our $\mathcal{F}_{\textsf{expn}}$ takes less communication traffic than that in SIRNN~\cite{rathee2021sirnn} with $s>12$. For $\mathcal{F}_{\textsf{sigmoid}}$ and $\mathcal{F}_{\textsf{Tanh}}$, we reduce the communication traffic by using fewer building blocks and our $\mathcal{F}_{\textsf{expn}}$ (when $s>12$). Fig.~\ref{fig:SigmoidComm} and \ref{fig:TanhComm} demonstrate the effectiveness of our strategy. Table~\ref{tab:Commloads} exhibits the communication traffic for one instance with $s=16$. It is shown that our implementation has much less communication traffic  than MP-SPDZ~\cite{keller2020mp}. Compared to SIRNN~\cite{rathee2021sirnn}, the communication traffic is reduced by more than $35\%$ for $\mathcal{F}_{\textsf{expn}}$, and reduced by more than $40\%$ for $\mathcal{F}_{\textsf{sigmoid}}$ and $\mathcal{F}_{\textsf{Tanh}}$.

\begin{table}[!t]
\caption{The communication traffic for one instance with $s$ = 16.}
\label{tab:Commloads}
\centering
\setlength{\tabcolsep}{3.1mm}{
\begin{tabular}{|c|cccc|}
\hline
\multicolumn{1}{|c|}{\multirow{2}{*}{\textbf{Technique}}}
& \multicolumn{4}{c|}{\textbf{Communication traffic } \textsf{(in KB)}}
\\ \cline{2-5}
\multicolumn{1}{|c|}{}
& \multicolumn{1}{c|}{$\mathcal{F}_{\textsf{exp}}$}
& \multicolumn{1}{c|}{$\mathcal{F}_{\textsf{expn}}$}
& \multicolumn{1}{c|}{$\mathcal{F}_{\textsf{sigmoid}}$}
& \multicolumn{1}{c|}{$\mathcal{F}_{\textsf{Tanh}}$}
\\ \hline
\textsf{MP-SPDZ~\cite{keller2020mp}}
& \multicolumn{1}{c|}{182.27}
& \multicolumn{1}{c|}{182.27}
& \multicolumn{1}{c|}{3030.36}
& - \\ \hline
\textsf{SIRNN~\cite{rathee2021sirnn}}
& \multicolumn{1}{c|}{-}
& \multicolumn{1}{c|}{3.80}
& \multicolumn{1}{c|}{7.47}
& 7.58 \\ \hline
\textsf{Ours}
& \multicolumn{1}{c|}{3.74}
& \multicolumn{1}{c|}{2.45}
& \multicolumn{1}{c|}{4.47}
& 4.50 \\ \hline
\end{tabular}
}
\end{table}

\subsection{Run-time}\label{sec:Runningtime}
\begin{table}[!t]
\caption{Run-time of the protocols for 16384 instances in LAN.}
\label{tab:RunTimesL}
\centering
\setlength{\tabcolsep}{3.2mm}{
\begin{tabular}{|c|cccc|}
\hline
\multicolumn{1}{|c|}{\multirow{2}{*}{\textbf{Technique}}}
& \multicolumn{4}{c|}{\textbf{Run-time} \textsf{(in ms)}}
\\ \cline{2-5}
\multicolumn{1}{|c|}{}
& \multicolumn{1}{c|}{$\mathcal{F}_{\textsf{exp}}$}
& \multicolumn{1}{c|}{$\mathcal{F}_{\textsf{expn}}$}
& \multicolumn{1}{c|}{$\mathcal{F}_{\textsf{sigmoid}}$}
& \multicolumn{1}{c|}{$\mathcal{F}_{\textsf{Tanh}}$}
\\ \hline
\textsf{MP-SPDZ~\cite{keller2020mp}}
& \multicolumn{1}{c|}{419}
& \multicolumn{1}{c|}{419}
& \multicolumn{1}{c|}{103621}
& - \\ \hline
\textsf{SIRNN~\cite{rathee2021sirnn}}
& \multicolumn{1}{c|}{-}
& \multicolumn{1}{c|}{172}
& \multicolumn{1}{c|}{314}
& 327 \\ \hline \multirow{2}{*}{\textsf{Ours}}
& \multicolumn{1}{c|}{143}
& \multicolumn{1}{c|}{95}
& \multicolumn{1}{c|}{186}
& 189 \\
& \multicolumn{1}{c|}{-}
& \multicolumn{1}{c|}{$(55\%)$}
& \multicolumn{1}{c|}{$(59\%)$}
& \multicolumn{1}{c|}{$(58\%)$}
\\ \hline
\end{tabular}
}
\end{table}

\begin{table}[!t]
\caption{Run-time of the protocols for 16384 instances in WAN.}
\label{tab:RunTimesW}
\centering
\setlength{\tabcolsep}{3.2mm}{
\begin{tabular}{|c|cccc|}
\hline
\multicolumn{1}{|c|}{\multirow{2}{*}{\textbf{Technique}}}
& \multicolumn{4}{c|}{\textbf{Run-time} \textsf{(in seconds)}}
\\ \cline{2-5}
\multicolumn{1}{|c|}{}
& \multicolumn{1}{c|}{$\mathcal{F}_{\textsf{exp}}$}
& \multicolumn{1}{c|}{$\mathcal{F}_{\textsf{expn}}$}
& \multicolumn{1}{c|}{$\mathcal{F}_{\textsf{sigmoid}}$}
& \multicolumn{1}{c|}{$\mathcal{F}_{\textsf{Tanh}}$}
\\ \hline
\textsf{MP-SPDZ~\cite{keller2020mp}}
& \multicolumn{1}{c|}{6.79}
& \multicolumn{1}{c|}{6.79}
& \multicolumn{1}{c|}{1884.15}
& - \\ \hline
\textsf{SIRNN~\cite{rathee2021sirnn}}
& \multicolumn{1}{c|}{-}
& \multicolumn{1}{c|}{3.16}
& \multicolumn{1}{c|}{5.71}
& 5.39 \\ \hline
\multirow{2}{*}{\textsf{Ours}}
& \multicolumn{1}{c|}{1.99}
& \multicolumn{1}{c|}{1.35}
& \multicolumn{1}{c|}{3.20}
& 3.15 \\
& \multicolumn{1}{c|}{-}
& \multicolumn{1}{c|}{$(43\%)$}
& \multicolumn{1}{c|}{$(56\%)$}
& \multicolumn{1}{c|}{$(58\%)$}
\\ \hline
\end{tabular}
}
\end{table}

 In this section, we test the run-time of the protocols. We pack the 16384 instances together with $s=16$, and record the total running time in both LAN and WAN environments. The results are obtained by 4 threads in parallel as listed in Table~\ref{tab:RunTimesL} and Table~\ref{tab:RunTimesW}. Compared with MP-SPDZ~\cite{keller2020mp}, our realization achieves two orders of magnitude faster. Compared with SIRNN~\cite{rathee2021sirnn}, the proposed $\mathcal{F}_{\textsf{expn}}$ runs about three times faster due to fewer communication rounds and less communication traffic. For $\mathcal{F}_{\textsf{sigmoid}}$ and $\mathcal{F}_{\textsf{Tanh}}$, our implementation also saves about $40\%$ of the running time compared to SIRNN.

\subsection{RNN  Inference Case Study}\label{sec:RNN}
\begin{table}[!t]
\caption{Inference experiments on Google-30 dataset with FastGRNN network (only \textsf{sigmoid} and \textsf{Tanh}).}
\label{tab:RNN}
\centering
\setlength{\tabcolsep}{0.2mm}{
\begin{tabular}{|c|c|c|c|}
\hline
\textbf{Technique} & \textbf{Communication traffic } & \textbf{LAN}\textsf{(in seconds)} & \textbf{WAN}\textsf{(in seconds)}
\\ \hline
\textsf{SIRNN}&98.93MB &10.17 &442.96
\\ \cline{1-4}
\multirow{2}{*}{\textsf{Ours}}&78.16MB &8.27  &340.06\\
&($79\%)$& ($81\%$)& ($77\%$)
\\ \hline
\end{tabular}
}
\end{table}

 In this section, we reproduce the inference experiments in~\cite{rathee2021sirnn}, where FastGRNN~\cite{kusupati2018fastgrnn} network is used to inference on Google-30 dataset. FastGRNN invokes 100 $\mathcal{F}_{\textsf{sigmoid}}$ and 100 $\mathcal{F}_{\textsf{Tanh}}$. For $\mathcal{F}_{\textsf{sigmoid}}$, the bitwidth for both input and output is 16, and the scale $s$ for input and output are 8 and 14, respectively. For $\mathcal{F}_{\textsf{Tanh}}$, the bitwidth for both input and output is also 16, and the scale $s$ for both input and output are 9. As shown in Table~\ref{tab:RNN}, our implementation saves about $20\%$ of time at the sigmoid and Tanh parts in LAN environment, and the saving is more in the WAN.

\section{Conclusion}

 In this paper, we improved the secure implementations for four non-linear functions in SIRNN, including exponential function, exponential function with nonpositive inputs, sigmoid function, and tanh function. We take advantage of the intrinsic features of functions as well as tiny tricks. We tested the precision, number of building blocks, communication rounds, communication traffic , and run-time of the proposed protocols, and the experimental results demonstrate our design outperforms the state-of-the-arts. We also test the performance of our protocols in the typical FastGRNN networks on Google-30 dataset. The results prove the correctness and efficiency of our protocols. Our strategies and contributions are not limited to the functionalities and applications presented in this paper. Our strategies have interesting effects on other symmetric functions as well, and our exponential protocol is not limited to RNN inference applications. It is also an indispensable building block for softmax activation functions commonly used in DNN training and Poisson regression.


\bibliographystyle{IEEEtran}
\bibliography{REF}

\begin{IEEEbiography}[{\includegraphics[width=1in,height=1.25in,clip,keepaspectratio]{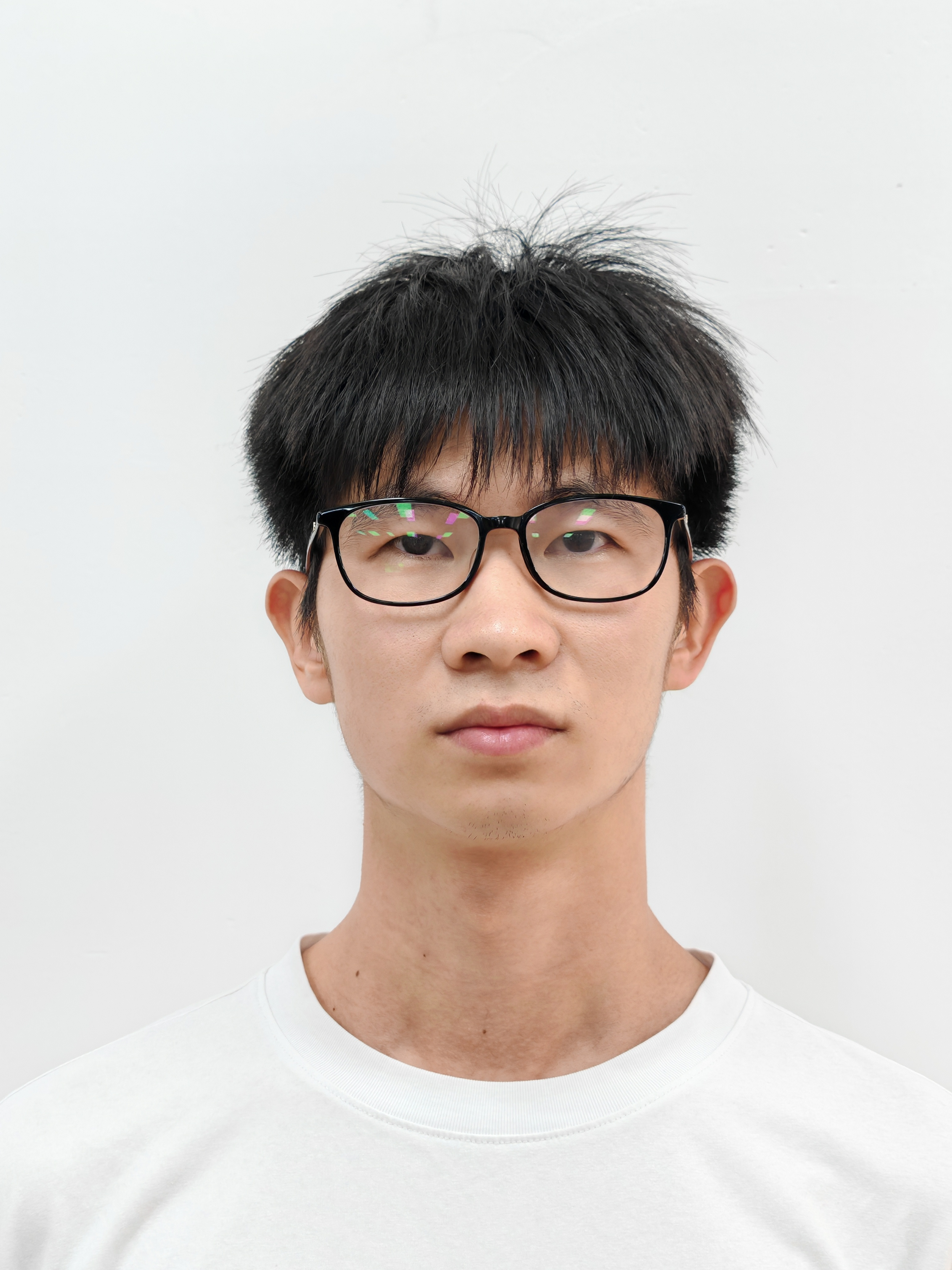}}]{Qian Feng}
received the B.S degree in Hunan University of Technology, Zhuzhou, China, in June 2021. He is currently pursuing the master's degree with the College of Cyberspace Security, Jinan University, Guangzhou, China. His research interests include Secure Multi-Party Computation and Privacy-preserving Machine Learning.
\end{IEEEbiography}

\begin{IEEEbiography}[{\includegraphics[width=1in,height=1.25in,clip,keepaspectratio]{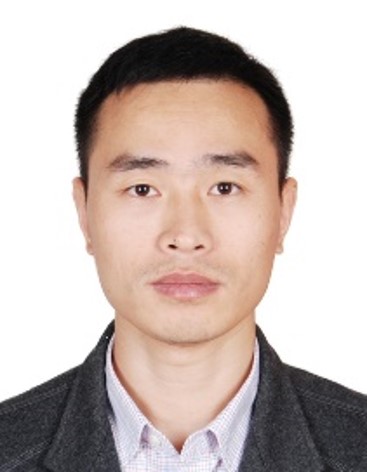}}]{Zhihua Xia }
(Member, IEEE) received the Ph.D. degree in computer science and technology from Hunan University, China, in 2011. He was a Visiting Scholar with the New Jersey Institute of Technology, USA, in 2015. He was a Visiting Professor with Sungkyunkwan University, South Korea, in 2016. He is currently a Professor with the College of Cyberspace Security, Jinan University, China. His research interests include AI security, cloud computing security, and digital forensics. He is also the Managing Editor
for IJAACS.
\end{IEEEbiography}

\begin{IEEEbiography}[{\includegraphics[width=1in,height=1.25in,clip,keepaspectratio]{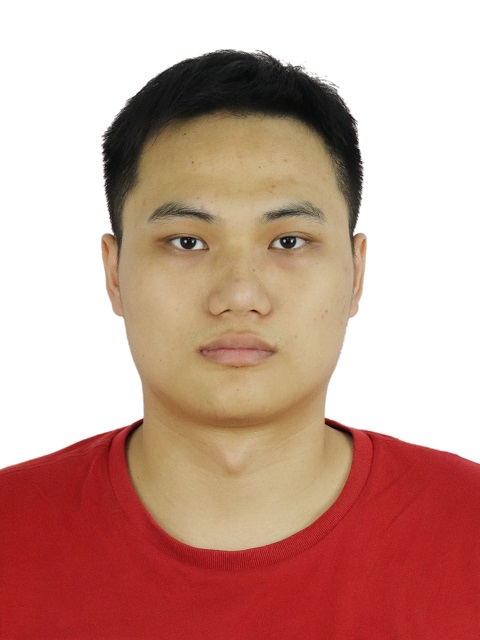}}]{Zhifeng Xu }
received the B.S degree in Dongguan University of Technology, Dongguan, China, in June 2021. He is currently pursuing the master's degree with the College of Cyberspace Security, Jinan University,Guangzhou, China. His research interests include Distributed Computing and Secure Multi-Party Computation.
\end{IEEEbiography}

\begin{IEEEbiography}[{\includegraphics[width=1in,height=1.25in,clip,keepaspectratio]{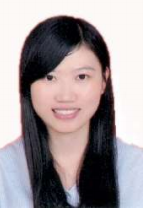}}]{Jiasi Weng}
obtained her doctoral degree from Jinan University in China and served as a postdoctoral researcher at City University of Hong Kong. She is currently an associate professor at Jinan University. In the past five years, she has published more than 10 high-level papers in internationally renowned and top-tier journals and conferences such as IEEE Transactions on Information Forensics and Security, IEEE Transactions on Dependable and Secure Computing, and IEEE INFOCOM.
\end{IEEEbiography}

\begin{IEEEbiography}[{\includegraphics[width=1in,height=1.25in,clip,keepaspectratio]{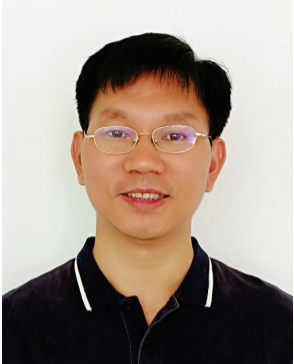}}]{Jian Weng}
	(Member, IEEE) received the Ph.D. degree in computer science and engineering from Shanghai Jiao Tong University, Shanghai, China, in 2008. He is currently a Professor and the Dean with the College of Information Science and Technology, Jinan University, Guangzhou, China. His research interests include public key cryptography, cloud security, and block-chain. He was the PC Co-Chairs or PC Member for more than 30 international conferences. He also serves as an Associate Editor for the IEEE Transactions on Vehicular Technology.
\end{IEEEbiography}

\vfill

\end{document}